\documentclass[aps,twocolumn,showpacs,showkeys,nofootinbib]{revtex4}
\usepackage{epsfig}
\usepackage{amsmath}
\usepackage{amsfonts}
\usepackage{amssymb}
\usepackage{graphicx}
\usepackage{colordvi}
\usepackage{subfigure}
\newcommand{\be}{\begin{equation}}
\newcommand{\ee}{\end{equation}}

\begin{document}

\title{Quantum superpositions of a mirror for experimental tests for nonunitary Newtonian gravity}

\author{Filippo Maimone\footnote{e-mail address:
\textit{f\_maimone@alice.it}}, Giovanni Scelza\footnote{e-mail
address: \textit{lucasce73@gmail.com}}, Adele
Naddeo$^{(1)}$\footnote{e-mail address:
\textit{naddeo@sa.infn.it}}, Vinicio Pelino }

\affiliation{(1)Dipartimento di Fisica ``E. R. Caianiello'',
Universit$\grave{a}$ degli Studi di Salerno and CNISM,
Unit$\grave{a}$ di Ricerca di Salerno, Via Ponte don Melillo, $
84084$ Fisciano (SA), Italy}

\begin{abstract}
Aim of this work is to calculate explicitly the result of the
experiment of superposition of a mirror in the Michelson photon
cavities interferometric device proposed by Marshall, Simon,
Penrose and Bownmeester, as expected within a recently proposed
model of non-unitary self-gravity inducing localization. As for
other proposals of modifications of Quantum Mechanics in a
non-unitary sense, aimed to account for both unitary evolution and
irreversible collapse, like in the famous Ghirardi-Rimini-Weber
and Pearle's models, it turns out that, for the experimental
parameters proposed, no effect is detectable at all. It is pointed
out that the enhancing properties of matter granularity does not
substantially change this conclusion. Parameters have also been
exploratively varied in a certain range beyond the proposed
values. It is shown that within `sensible' parameters, that are
not yet attainable within current technology, the model exhibits a
peculiar signature with respect to other collapse models as far as
parameters space is explored. Besides, the calculation offers a
way to see non-unitary gravity at work in a quasi-realistic
setting.
\end{abstract}

 \maketitle



\section{Introduction}

In the past few decades several proposals of modification of
Quantum Mechanics (QM) have appeared in the literature, aimed at
unifying its internal fundamental dichotomy between unitary
deterministic quantum dynamics and non-linear irreversible state
collapse following a measurement process \cite{17}. On the other
hand big efforts have been devoted towards an attempt to reconcile
Einstein gravity with quantum theory. In this context, some
approaches have focused on the possible role of gravity in state
function collapse as a result of the incompatibility of general
relativity and the unitary time evolution of QM \cite{0,13,14,10}.
It has been shown, in fact, that the existence of linear
superpositions of states with macroscopic mass-distribution
differences would entail a breakdown of classical space-time
making the traditional quantum dynamics somehow troubling
\cite{0}. As distinct from Penrose proposal, some other detailed
collapse models have been proposed, which are based on a
spontaneous stochastic state vector reduction: the
Ghirardi-Rimini-Weber (GRW) model \cite{grw}, the quantum
mechanics with universal position location model (QMUPL) of wave
function collapse \cite{13} and the continuous spontaneous
localization (CSL) model \cite{csl}. Recently the mechanism of
spontaneous symmetry breakdown of time translation symmetry has
been invoked as well in order to give rise to the quantum state
reduction \cite{wezel}.

In a different proposal, De Filippo introduced a nonunitary model
of Newtonian gravity (NNG, from now on), which can be seen as the
non-relativistic limit of a classically stable version of higher
derivative gravity (see e.g. Ref.\cite{1}, references therein).
This model presents several appealing features to become a natural
candidate for an effective low-energy model of gravity. For
example, while reproducing at a macroscopic level the ordinary
Newtonian interaction, it presents a mass threshold for
gravitational localization, which for ordinary matter densities is around $%
10^{11}$ proton masses \cite{15}. The model can be seen as a
realistic version of the nonunitary toy models \cite{2,3,4}
inspired by the emergence of the information loss paradox
\cite{5,6,7} from black hole physics. On the other hand the
violation of unitarity when matching quantum mechanics and gravity
was argued also outside black hole physics, on general consistency
grounds \cite{0,8}. The model affords a mechanism for the
evolution of macroscopic coherent superpositions of states into
ensembles of pure states, each one of them corresponding -- within
a future consistent general covariant theory -- to an unambiguous
space-time. Its features include its ability to produce an
evolution of the density matrix compatible with the expectations
leading to the phenomenological spontaneous localization models,
as it was argued that they should be both nonlinear and nonunitary
\cite{10}. While sharing with the other proposals the non-linear
non-unitary character, at variance with them, however, it does not
present obstructions consistent with its special-relativistic
extension \cite{11}. Another success of the model is the emergence
of a unified picture for ordinary and black hole entropy as
entanglement entropy with hidden degrees of freedom \cite{1}, in
agreement with Bekenstein-Hawking entropy \cite{bh} and Hawking
evaporation temperature; that arises from the smoothed singularity
of the black hole introduced by the model and paves the way for
the quantum foundations of the second law of thermodynamics.

It is important to realize that the subject of a fundamental
non-unitarity, and the various detailed mechanisms proposed
\cite{0,8,12,13,14,15bis,16}, is not just a matter of philosophy,
but could be, in principle, experimentally proved or disproved.
Current technological progresses which have being achieved in
isolating, manipulating and controlling an higher and higher
number of degrees of freedom indicate a not far possibility of
detecting fundamental decoherence, which would manifest in a clean
way only once the system has been sufficiently protected against
the sources of environmental noise \cite{18}. Indeed an experiment
designed to detect fundamental deviations from unitary quantum
evolution would be of considerable importance. Some technologies
and devices have, at present, been recognized to be particularly
suitable to create quantum state superpositions which are
\textit{macroscopically} distinct \cite{leggett}. Among them,
there are diffraction of complex molecules up to $2\times 10^{3}$
proton masses \cite{exp1}, current cat states in SQUID devices
\cite{exp2}
and superpositions of atomic matter waves in Bose Einstein condensates \cite%
{exp3}.

Recent progress in optomechanical systems may soon allow one to
make superpositions of even larger objects, such as micro-sized
mirrors or cantilevers \cite{exp4}, and to test quantum phenomena
at larger scales. In this context an appealing quite recent
proposal for the practical realization of the Penrose ``gedanken
experiment'' \cite{0} considers the relatively small
\textit{CM}-displacement of a lump of $10^{14}$ proton masses in a
interferometric device in which two high-finesse optical cavities
are inserted into its arms \cite{29}. The cavity in arm \textbf{A}
has a very small end mirror mounted on a micro-mechanical
oscillator (cantilever), which suffers the radiation pressure of
the photon confined inside it and as a consequence can be excited
into a distinguishable quantum state. A single photon incident on
a 50-50 beam splitter will realize a superposition of being in
either of the two arms; then, the coupling between the photon and
the cantilever will lead to an entangled state putting the
cantilever into a superposition of distinct positions. After a
full mechanical period of the cantilever, it recovers its original
position; if the photon leaks out of the cavity at this stage, a
revival of the interference (visibility) is observed, provided
that the quantum superposition state of the system survives at the
intermediate times. Conversely, if the state of the system
collapses due to some decoherence mechanism, visibility will not
revive. Summarizing, a measurement of the magnitude of the revival
of visibility gives a measurement of decoherence occurred in the
time interval under consideration.

Our work is devoted to calculate explicitly the output of this
experiment \cite{29} according to nonunitary Newtonian gravity
model \cite{1,15,23,24}. We would point out that detailed
calculations for the expectations of some other collapse models,
gravitational or not, have already been done,
demonstrating a far reaching possibility to confirm the theory \cite%
{31,32,33}.

The plan of the paper is as follows. Section II contains a brief
description of the basic NNG model. In Section III we give a
qualitative discussion to argue, in the relevant parameter space
of the experiment by Marshall et al. \cite{29}, the subspaces
where gravitational effect could, in principle, be visible with
both homogeneous and granular assumption on mirror mass
distribution. Section IV is devoted to the application of NNG
model to the experiment, and its general solution. Then, in
Section V, we use the Wigner function to monitor the mirror's
state, and verify its behavior in time, after a measurement of
photons' state. Finally, in Sec.VI, we draw conclusions and
outline perspectives of this work. Calculational details are
devoted to Appendices.

\section{Non-unitary Newtonian Gravity model}
\setcounter{equation}{0} The aim of this Section is to briefly recall the key features of the
NNG model which we will use for calculations. On the basis of a number of considerations (among which consistency with basic formal
relations of QM, approximate energy conservation, classical and
quantum behavior of matter, requirement that non-unitary terms
have a gravitational origin, etc), it is possible to isolate a
two-parameter class of non-unitary gravity models, as discussed in
detail in Refs. \cite{1,15,23,24}. We will comment later in the Section on
these parameters, while we give here a concise definition of the
model in its simplest form, which will allow us to carry out our
calculations.

Let $H[\psi^{\dagger },\psi ]$ be the non-relativistic Hamiltonian
of a finite number of particle species, like electrons, nuclei,
ions, atoms and/or molecules, where $\psi ^{\dagger },\psi $ denote
the whole set $\psi _{j}^{\dagger }(x),\psi _{j}(x)$ of
creation-annihilation operators, \textit{i.e.} one couple per
particle species and spin component. $H[\psi ^{\dagger },\psi ]$\
includes the usual electromagnetic interactions accounted for in
atomic, molecular and condensed-matter physics. To incorporate
that part of gravitational interactions responsible for
non-unitarity, one has to introduce complementary
creation-annihilation operators $\widetilde{\psi }_{j}^{\dagger
}(x),\widetilde{\psi }_{j}(x)$ and the overall (meta-)Hamiltonian:
\begin{equation}
\begin{split}
H_{tot}&=H[\psi ^{\dagger },\psi ]+H[\widetilde{\psi }^{\dagger
},\widetilde{\psi }]+\\
&-{G}\sum_{j,k}m_{j}m_{k}\int d\textbf{x}d\textbf{y}\frac{\psi
_{j}^{\dagger }(\textbf{x})\psi _{j}(\textbf{x})\widetilde{\psi
}_{k}^{\dagger }(\textbf{y})\widetilde{\psi}_{k}(\textbf{y})}{
|\textbf{x}-\textbf{y}|},
\end{split}
\label{meta-hamiltonian}
\end{equation}
acting on the product $F_{\psi }\otimes F_{\widetilde{\psi }}$ of
Fock spaces of the $\psi $ and $\widetilde{\psi }$ operators,
where $m_{i}$ is the mass of the $i$-th particle species and $G$
is the gravitational constant. The $\widetilde{\psi }$ operators
obey the same statistics as the corresponding operators $\psi $,
while $[\psi ,\widetilde{\psi }]_{-}=[\psi , \widetilde{\psi
}^{\dagger }]_{-}=0$.

The meta-particle state space $S$ is the subspace of $F_{\psi
}\otimes F_{ \widetilde{\psi }}$, including the meta-states
obtained from the vacuum $ \left\vert \left\vert 0\right\rangle
\right\rangle =\left\vert 0\right\rangle _{\psi }\otimes
\left\vert 0\right\rangle _{\widetilde{\psi } } $ by applying
operators built in terms of the products $\psi _{j}^{\dagger
}(x)\widetilde{\psi }_{j}^{\dagger }(y)$ and symmetrical with
respect to the interchange $\psi ^{\dagger }\leftrightarrow
\widetilde{\psi }^{\dagger }$; as a consequence they have the same
number of $\psi $ (physical) and $\widetilde{\psi }$ (hidden)
meta-particles of each species. Since constrained meta-states
cannot distinguish between physical and hidden operators, the
observable algebra is identified with the physical operator
algebra. In view of this, expectation values can be evaluated by
preliminarily tracing out the $\widetilde{\psi }$ operators. In
particular, the most general meta-state corresponding to one
particle states is represented by
\begin{equation}
\left\vert \left\vert f\right\rangle \right\rangle =\int
d\textbf{x}\int d\textbf{y}\,f(\textbf{x},\textbf{y})\psi
_{j}^{\dagger }(\textbf{x})\widetilde{\psi }_{j}^{\dagger
}(\textbf{y})\left\vert \left\vert 0\right\rangle \right\rangle. \label{f}
\end{equation}
$$f(\textbf{x},\textbf{y})=f(\textbf{y},\textbf{x})$$
This is a consistent definition since $H_{tot}$\ generates a group of
(unitary) endomorphisms of $S$.

If we prepare a pure $n$-particle state, represented in the
original setting,
excluding gravitational interactions, by%
\begin{equation*}
\left\vert g\right\rangle =\int d^{n}\textbf{x}\,g\left(
\textbf{x}_{1},\textbf{x}_{2},...,\textbf{x}_{n}\right) \psi
_{j_{1}}^{\dagger }(\textbf{x}_{1})\psi _{j_{2}}^{\dagger
}(\textbf{x}_{2})...\psi _{j_{n}}^{\dagger
}(\textbf{x}_{n})\left\vert 0\right\rangle ,
\end{equation*}
its representation in $S$ is given by the metastate
\begin{eqnarray*}
&&\int d^{n}\textbf{x}\ d^{n}\textbf{y}\biggl(g\left(
\textbf{x}_{1},...,\textbf{x}_{n}\right) g\left(
\textbf{y}_{1},...,\textbf{y}_{n}\right) \times\\
&\times&\psi _{j_{1}}^{\dagger }(\textbf{x}_{1})...\psi
_{j_{n}}^{\dagger }(\textbf{x}_{n})\ \widetilde{\psi }_{j_{1}}^{\dagger }(\textbf{y}_{1})...%
\widetilde{\psi }_{j_{n}}^{\dagger }(\textbf{y}_{n})\left\vert
\left\vert 0\right\rangle \right\rangle\biggr) .
\end{eqnarray*}

A comment is in order on the possible extensions of the model
outlined in  Refs.\cite{1,15,23,24}. As said in the beginning of
the Section, the phenomenological general model depends on the two
parameters $(N,\varepsilon )$. The first refers to the number $N$
of copies in interaction, which on thermodynamical grounds can be
inferred to be $2$ (as in the model presented above). It is
however interesting to note that the limit $N\rightarrow \infty $
(with $\varepsilon =1$) reproduces the famous non-linear
Newton-Schr\"{o}dinger equation, sometimes considered in the
literature as a possible candidate equation for the inclusion of
the self-gravity in QM, relevant to the quantum-classical
transition \cite{15,10}. The second, $\varepsilon,$ modulates the
degree of nununitarity encoded in the gravitational interaction.
The above model definition corresponds to $\varepsilon =1$, for
which all Newtonian interaction is of nonunitary type. This choice
has been made in order to maximize the effect of nonunitarity,
while calculating the prediction on the experiment by Marshall et
al. \cite{29} in the best model-setting which gives the largest
possible deviations from unitarity.
\section{A semi-quantitative argument for NNG effects in mirror experiment}
\setcounter{equation}{0} Before considering the detailed
application to the mirror experiment of Ref. \cite{29}, in this
Section we give a semi-quantitative argument for a gross
identification of NNG effects.

When considering self-interaction gravitational energy, the
threshold mass of localization, $M_{tr}$ $\sim 10^{11}$proton\
masses $(=1.672\times 10^{-16}Kg) $, can be identified in the
following way. If $M<M_{tr}$ the metasystem behaves like an
hydrogen-like system, while in the case $M>M_{tr}$ the hidden
mass-copy is quite well superposed to the physical one. As a
consequence, the interaction potential can be approximated, within
the lowest energy state of the meta-system, by the harmonic
oscillator ground state with gaussian wave function width
\begin{equation*}
\Lambda _{G}=\left( \frac{\hbar }{\sqrt{\left( 4/3\right) \ \pi G\
\rho _{sil}\ M^{2}}}\right) ^{1/2},
\end{equation*}
where $\rho _{sil}=5\times 10^{3}$ $Kg/m^{3}$ is silicon density.
For nonunitary gravity to be effective in localizing the mirror,
this length scale must be at least comparable with the wave
packets separation:
\begin{equation}
\Delta x=\kappa \sqrt{\frac{\hbar }{2M\omega _{m}}},  \label{dx}
\end{equation}%
where $\omega _{m}$ is the mirror's frequency and $\kappa$ is the
optomechanical coupling constant; then the condition $\Delta
x\gtrsim \ \Lambda _{G}$ amounts to
\begin{equation}
\varkappa \equiv \ \frac{1}{\rho _{sil}\ G\ }\left( \frac{\omega _{m}}{%
\kappa ^{2}}\right) ^{2}\lesssim 1;  \label{eqn:ineq}
\end{equation}
for the experimental parameters,\textit{ i.e.} $\omega _{m}\simeq~
2\pi \times 500 Hz,\ \rho _{sil}\simeq 5\times 10^{3}Kg/m^3$ and
$\ \kappa \sim 1$, we get $\Delta x\simeq 5.79\times 10^{-14}m$,
$\varkappa \sim 10^{13}$. As pointed out in Ref
{\cite{Diosi_granulare}}, an enhancement in the possibility of
observing gravitational decoherence effects is provided by taking
into account the real distribution of mass inside a crystal, which
is very concentrated within nuclei (see Appendix A.2). In that
case, one should consider instead of $\rho _{sil}$ a matter
density $\rho _{nuc}\sim 10^{4}\rho _{sil}$, given by a silicon
nucleus mass divided by its effective radius, which can be
estimated as the typical spread of the wavefunction inside a
crystal. This leads to $\varkappa \sim 10^{9}$, which is still
much greater than unity.

It should be stressed that the choice between homogeneous or
granular matter distribution is not arbitrary, but is dictated by
the experimental situation. As a matter of fact, when the relative
displacement of meta-masses $\Delta x$ is of the order of the
nucleus effective radius, it seems appropriate to take into
account granularity; when  $\Delta x$ is made much greater than
interatomic separation then homogeneity assumption appears to be
the most suitable one; finally, for $\Delta x$ of the order of
interatomic separation, if imperfections like dislocations are
present in the sample (as it usually happens, even when very
accurate preparation methods are used), then meta-masses are
likely to `feel' an effective homogeneous masses potential (this
is because in the presence of a sufficient number of dislocations,
as two meta-nuclei get nearer and nearer in one place, two nuclei
in another place in the crystal can equally well go farther and
farther from each other); otherwise, for a really perfect crystal
granularity should come again into play.

On the other hand, (fundamental) decoherence rate must be at least
comparable with (or lower than) a period of natural oscillation of
the mirror, which, in its turn, must be comparable with (or lower
than) environmental decoherence rate for the experiment to be
feasible.

Then the following chain of relations must be satisfied:
\begin{equation}
\frac{E_{G}}{\hbar }\sim \frac{\ \pi \kappa ^{2}\hbar G\rho
_{sil}\left( \rho _{nuc}\right) }{3\omega _{m}}\sim \omega
_{m}\gtrsim \gamma _{D}, \label{ineq2}
\end{equation}
where $E_{G}$ is the gravitational interaction energy of the
meta-masses (see Appendix A), $\gamma _{D}$ is the environmental
decoherence rate of the mirror \cite{29,31}. With the parameters
of the experiment the value $\omega _{m}=\omega _{m}^{\exp }\simeq
2\pi \times 500Hz$ has been proposed.

An exploration of parameter space within the exact solution has
confirmed
that first inequality (\ref{eqn:ineq}) and approximate equality in Eq. (\ref%
{ineq2}) must hold in order to see some relevant deviation from
unitarity.

It should be stressed that, in spite of the improvement of mass
size in the mirror experiment with respect to double-slit type
experiments, the degree of \textit{macroscopicity} of
superposition is controlled not only by mass but also by space
separation of the superposed wave packets.

A comment is in order on the apparent independence of the above
result on the mass. The main point is that we have chosen to
measure CM state separation in units of coherent states' size,
which means that a larger mass is associated with a smaller unit
of length and then, fixing $\kappa $, to a narrower peak
separation. One could also choose to fix the absolute separation
$\overline{\Delta x}$, and express $\kappa \equiv \kappa \left(
M,\omega _{m}\right) =\overline{\Delta x}\sqrt{\frac{2M\omega
_{m}}{\hbar }}$ in the above formulas.
\section{Application of NNG model to the mirror experiment}
\setcounter{equation}{0} In this Section we carry out interference
visibility calculations within the mirror experiment \cite{29} in
the framework of the NNG model \cite{1,15,23,24}.\\
To this aim, let us start by defining the gravity-free Hamiltonian
\begin{widetext}
$$H_{free}\left[b,b^{\dagger},\mathcal{N}_{A},\mathcal{N}_{B};\omega_{m}
\right]=\hbar\omega
_{Ph}\left(\mathcal{N}_{A}+\mathcal{N}_{B}\right)+\hbar\omega
_{m}b^{\dagger }b-\hbar g\mathcal{N}_{A}\left( b+b^{\dagger
}\right),$$
\end{widetext}
where $g=\kappa\,\omega _{m},$ $\mathcal{N}_{A,B}$ are the number
operators for the photon in the interferometer arms $A$ and $B$
respectively, while $b$ and $b^{\dag}$ are the phonon destruction
and creation operators associated with the motion of the mirror's
CM. In this way our (meta-)Hamiltonian can be written as
\begin{widetext}
$$H_{tot}=H_{free}\left[b,b^{\dagger
},\mathcal{N}_{A};\omega _{m}^{\ast }\right] +H_{free}\left[
\tilde{b},\tilde{b}^{\dagger },\widetilde{\mathcal{N }}_{A};\omega
_{m}^{\ast }\right] -K_{G}\left( b+b^{\dagger }\right)
(\widetilde{b}+\widetilde{b}^{\dagger}),$$
\end{widetext}
with $\omega _{m}^{\ast }=\omega _{m}\sqrt{1+2\frac{K_{G}}{\hbar
\omega _{m}}}$ (see Appendix A for the calculation of the
gravitational interaction strength $K_{G}$ in both homogeneous and
granular case). In practice, the (relevant degrees of freedom)
meta-system is formed by two gravitationally coupled harmonic
oscillators and two by two photonic modes, each couple of modes
interacting with its own mirror. Then, Schr\"{o}dinger state at
time \textit{t} is given by:
\begin{widetext}
\begin{equation}\label{eqn:prsch}
  \Vert \Psi (t)\rangle \rangle \equiv \Vert \Psi (t)\rangle \rangle
_{Sch}=\frac{1}{\pi^2}\iint d^{2}\beta d^{2}\widetilde{\beta
}\;{\Large K}^{\mathcal{N}_{A} \widetilde{\mathcal{N}}_{A}}(\beta
,\widetilde{\beta };t)|\psi (t)\rangle _{\beta }\otimes |\psi
(t)\rangle _{\widetilde{\beta }}\, ,
\end{equation}
\end{widetext}
where
\begin{eqnarray*}
&&\left\vert \psi \left( t\right) \right\rangle _{\beta }= \\
&=&\frac{1}{\sqrt{2}}e^{-i\omega _{Ph}t}\left( \left\vert
0_{A}1_{B}\right\rangle \otimes \left\vert \beta _c\right\rangle
+f(\beta )\left\vert 1_{A}0_{B}\right\rangle \otimes \left\vert
\beta _{l}\right\rangle \right),
\end{eqnarray*}
with
\begin{eqnarray*}
f(\beta ) &=&e^{i\kappa ^{2}\left( \omega _{m}^*t-\sin \omega
_{m}^*t\right)
}e^{i\kappa \Im \left[ \beta \left( 1-e^{-i\omega _{m}^*t}\right) \right] } \\
\left\vert \beta _{c}\right\rangle &=&\left\vert \beta e^{-i\omega
_{m}^*t}\right\rangle ;\left\vert \beta _{l}\right\rangle
=\left\vert \beta e^{-i\omega_{m}^*t}+\kappa \left( 1-e^{-i\omega
_{m}^*t}\right) \right\rangle.
\end{eqnarray*}
Here $\Im{(x)}$ denotes the imaginary part of $x$. Computational
details on kernel $K$ are devoted to Appendix B.

Since the only experimentally accessible quantity is the
\textit{visibility}, defined as twice the off-diagonal term (in
the absolute value) of the physical photon's state $\rho_{AB}$, we
are going to calculate this quantity as
$$Vis\left( t\right) =2\left\vert
Tr_{m,\widetilde{m}}Tr_{\widetilde{Ph}}R_{V}^{\left( \alpha
\right) }\right\vert ,$$
where $R_{V}^{\left( \alpha\right)}$ is
defined as
$$R_{V}^{\left( \alpha \right) }=\left\langle 1_{A}0_{B}\left\vert \left\vert
\Psi \right\rangle \right\rangle \left\langle \left\langle \Psi
\right\vert \right\vert 0_{A}1_{B}\right\rangle .$$

It can be shown that visibility has a one-to-one correspondence
with von Neumann entropy, which represents a good measure of
entanglement for a pure bipartite state \cite{31}.

We write the kernel as:
\begin{widetext}
\begin{eqnarray*}
{\Large K}^{\mathcal{N}_{A}\widetilde{\mathcal{N}}_{A}}(\beta
,\widetilde{\beta },t)&=&\underset{
\mathbb{K}^{\mathcal{N}_{A}\widetilde{\mathcal{N}}_{A}}\left(
t\right) }{\underbrace{\Lambda \left( t\right) e^{
-\frac{K_{G}}{2\hbar ^{2}}\biggl[ 4\alpha \left[
\mathcal{F}_{1}^{\gamma ^{\ast },\Gamma _{+}}+\alpha
\mathcal{F}_{1}^{\gamma ^{\ast },\gamma ^{\ast }}\right]
+\mathcal{F}_{1}^{\Gamma _{+},\Gamma _{+}}+
\mathcal{F}_{2}^{\Gamma _{-},\Gamma _{-}}\biggr]}\;e^{-|\alpha|^2}}}\times\\
&\times&
e^{-\frac{|\beta|^2}{2}-\frac{|\widetilde{\beta}|^2}{2}+\beta^*\alpha
+\widetilde{\beta}^*\alpha}
\times\\
&\times & e^{-\frac{K_{G}}{2\hbar ^{2} }\biggl[
\underset{F_{p}}{\underbrace{\left( \mathcal{F}_{1}^{\gamma
,\gamma }+\mathcal{F}_{2}^{\gamma ,\gamma }\right) }}\beta ^{\ast
2}+2\underset{ F_{\Gamma p}}{\underbrace{( \mathcal{F}_{1}^{\gamma
,\Gamma _{+}}+ \mathcal{F}_{2}^{\gamma ,\Gamma _{-}}+2\alpha
\mathcal{F}_{1}^{\gamma ^{\ast },\gamma
}) }}\beta ^{\ast }\biggr]}\times\\
&\times & e^{ -\frac{K_{G}}{2\hbar ^{2}}\biggl[
\underset{F_{p}}{\underbrace{ \left( \mathcal{F}_{1}^{\gamma
,\gamma }+\mathcal{F}_{2}^{\gamma ,\gamma }\right)
}}\widetilde{\beta }^{\ast 2}+2\underset{F_{m}}{\underbrace{\left(
\mathcal{F}_{1}^{\gamma ,\gamma }-\mathcal{F}_{2}^{\gamma ,\gamma
}\right) }} \beta ^{\ast }\widetilde{\beta }^{\ast
}+2\underset{F_{\Gamma m}}{ \underbrace{( \mathcal{F}_{1}^{\gamma
,\Gamma _{+}}-\mathcal{F} _{2}^{\gamma ,\Gamma _{-}}+2\alpha
\mathcal{F}_{1}^{\gamma ^{\ast },\gamma }) }}\widetilde{\beta
}^{\ast }\biggr]}
\end{eqnarray*}
\end{widetext}
where the functions $\mathcal{F}_{i}=\mathcal{F}_{i}(\omega_m^*t)$
are defined in Appendix B, Eq. (\ref{eqn:f1}). Then we proceed
to write the products ${\Large K}^{\mathcal{N}
_{A}\widetilde{\mathcal{N}}_{A}}(\beta ,\widetilde{\beta
},t){\Large K}^{ \mathcal{N}_{A}^{\prime
}\widetilde{\mathcal{N}}_{A}^{\prime }\ast }(\beta ^{\prime
},\widetilde{\beta }^{\prime },t)$ in the form ($k^{\prime
}=K_{G}/2\hbar ^{2}$):
\begin{widetext}
\begin{eqnarray*}
&&\mathbb{K}^{\mathcal{N}_{A}\widetilde{\mathcal{N}}_{A}}\left(
t\right) \mathbb{K}^{\mathcal{N}_{A}^{\prime
}\widetilde{\mathcal{N}}_{A}^{\prime } \ast }\left( t\right)
\underset{{\Large K}_{1}{\Large (\beta )}}{\underbrace{e^
{-k^{\prime }\left[ F_p\beta ^{\ast 2}+2F_{\Gamma p}\beta ^{\ast
}\right] }}}\ \underset{{\LARGE K }_{2}\left( \beta
,\widetilde{\beta }\right) }{\underbrace{e^{- k^{\prime } \left[
F_p\widetilde{\beta }^{\ast 2}+2F_{m}\beta ^{\ast }
\widetilde{\beta }^{\ast }+2F_{\Gamma m}\widetilde{\beta }^{\ast
}\right] }}} \underset{{\LARGE K}_{3}\left( \beta ^{\prime
}\right) }{\underbrace{e^{- k^{\prime }\left[ F^*_p\beta^{\prime
2}+2F_{\Gamma p}^*\beta ^{\prime }\right] }}}\ \underset{{\Large
K}_{4}\left( \beta ^{\prime }, \widetilde{\beta }^{\prime }\right)
}{\underbrace{e^{- k^{\prime }\left[ F^*_p\widetilde{\beta
}^{\prime 2}+2F_{m}^*\beta ^{\prime }\widetilde{ \beta }^{\prime
}+2F_{\Gamma m}^*\widetilde{\beta }^{\prime }\right] }}}\times\\
&\times&
\underbrace{e^{-\frac{|\beta|^2}{2}-\frac{|\beta'|^2}{2}+\beta^*\alpha+\beta'\alpha^*
}}_{h(\beta,\beta')}\underbrace{e^{-\frac{|\widetilde{\beta}|^2}{2}-\frac{|\widetilde{\beta'}|^2}{2}+\widetilde{\beta}^*\alpha
+\widetilde{\beta}'\alpha^*}}_{g(\widetilde{\beta},\widetilde{\beta}')}.
\end{eqnarray*}
Let's calculate the traces:
\begin{eqnarray*}
&&Tr_{m,\widetilde{m}}Tr_{\widetilde{Ph}}R_{V}^{\left( \alpha
\right) }= \frac{1}{4\pi^4}e^{i\kappa ^{2}\left( \omega _{m}^{\ast
}t-\sin \omega^* _{m}t\right) }\times\\
&\times &\iiiint
d^2(\beta,\beta',\widetilde{\beta},\widetilde{\beta}')\;e^{i\kappa\Im[\beta(1-e^{-i\omega_m^*t})]}
\underset{{\Large L} \left( \beta ,\beta ^{\prime }\right)
}{\underbrace{e^{\beta _{l}\beta _{c}^{\prime \ast
}-\frac{1}{2}\left\vert \beta' _{c}\right\vert ^{2}-\frac{1
}{2}\left\vert \beta _{l}\right\vert ^{2}}}}\biggl[ {\Large
K}^{10}(\beta ,\widetilde{\beta }){\Large K}^{00\ast }( \beta
^{\prime },\widetilde{\beta }^{\prime }) \underset{{\Large H}
_{c}\left( \widetilde{\beta },\widetilde{\beta }^{\prime }\right)
}{ \underbrace{e^{\widetilde{\beta} _{c}\widetilde{\beta}
_{c}^{\prime \ast }-\frac{1}{2}\left\vert \widetilde{\beta}
_{c}\right\vert ^{2}-\frac{1}{2}\left\vert \widetilde{\beta}
_{c}^{\prime }\right\vert ^{2}}}}\;+\\
&+&{\Large K}^{11}(\beta ,\widetilde{\beta }){\Large K}^{01\ast }(
\beta ^{\prime },\widetilde{\beta }^{\prime
})\underbrace{e^{i\kappa\Im[\widetilde{\beta}(1-e^{-i\omega_m^*t})]}}_{\varepsilon(\widetilde{\beta})}
\underbrace{e^{-i\kappa\Im[\widetilde{\beta}'(1-e^{-i\omega_m^*t})]}}_{\delta(\widetilde{\beta}')}
\underset{{\Large H} _{l}\left( \widetilde{\beta
},\widetilde{\beta }^{\prime }\right) }{
\underbrace{e^{\widetilde{\beta }_{l}\widetilde{\beta
}_{l}^{\prime \ast }- \frac{1}{2}\left\vert \widetilde{\beta
}_{l}\right\vert ^{2}-\frac{1}{2} \left\vert \widetilde{\beta
}_{l}^{\prime }\right\vert ^{2}}}} \biggr].
\end{eqnarray*}
\end{widetext}

Visibility is then given by two contributions:
\begin{equation}
Vis\left( t\right) =2\left\vert Tr_{m,\widetilde{m}}Tr_{\widetilde{Ph}%
}R_{V}^{\left( \alpha \right) }\right\vert =2\left\vert \left( I\right)
+\left( II\right) \right\vert .  \label{visibbility}
\end{equation}

Here we do not write the explicit integrals for $(I)$ and $(II)$,
which can be found in Appendix C; it is worth noting that, if we
discard their photons number dependence, they are formally
similar. In the free case of no NNG interaction, \textit{i.e.} by
putting $k^{\prime }=0$, and for $ \alpha =0$, we get
\begin{widetext}
\begin{eqnarray*}
(I)\equiv(II)&=&\frac{1}{4}|\Lambda(t)|^2e^{i\kappa ^{2}\left(
\omega _{m}t-\sin \omega _{m}t\right)}
\,e^{-\kappa^2(1-\cos\omega_mt)}\qquad \bigl(\mbox{for
}k'=0,\mbox{ }\omega_m^*=\omega_m\mbox{ and
}|\Lambda(t)|^2=1\bigr)
\end{eqnarray*}
\end{widetext}as found in \cite{29}.
\begin{figure}[tbph]
\centering
\includegraphics[height=40mm]{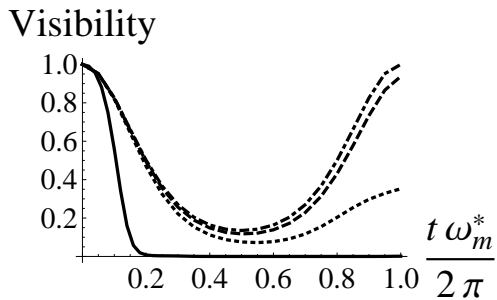}
\caption{Interference visibility as a function of time in the homogeneous case plotted
for $\protect\kappa =1,$ $ mass=5\times 10^{-12}kg$ for the mirror
size $L=10^{-5}m$ and several values of the frequency: $ \omega
_{m}=\omega _{m}^{exp}\times 10^{-5}$ (dot-dashed),   $ \omega
_{m}=\omega _{m}^{exp}\times 10^{-6}$ dashed,  $ \omega
_{m}=5\,\omega _{m}^{exp}\times 10^{-7}$ (dotted) and $ \omega
_{m}=\omega _{m}^{exp}\times 10^{-7}$ (continuous black line). Visibility and time are expressed in dimensionless units.}
\label{fig:Visibility}
\end{figure}

The behavior of visibility for $\kappa =1$ and four different
values of $\omega _{m}$ is depicted in Figure
\ref{fig:Visibility}. The first case shows no difference with the
free case, and the remaining three, in decreasing order,
progressively give a more and more reduction of revival effect at
the end of the cycle. A direct comparison can be made with the
predictions of other phenomenological collapse models, like GRW,
QMUPL and CLS \cite{10,32}. Let us focus, for example, on the GRW
model \cite{grw}. In such a case the explicit formula for
visibility is
\begin{equation*}
Vis\left( t\right) =Vis_{0}\left( t\right) \ e^{-\frac{3\ \kappa
^{2}\hbar \ \eta _{0}}{2\ \mu_{nuc}\,\omega_{m}}\left(
t-\frac{4}{3}\frac{\sin \omega_{m}t}{\omega_{m}}+\frac{\sin
2\omega_{m}t}{6\,\omega_{m}}\right) },
\end{equation*}
where $Vis_{0}$ is the visibility in the free case, \textit{i.e.}
in the absence of any mechanism of decoherence (of fundamental or
environmental nature), $ \eta _{0}\simeq 0.5\times
10^{-2}s^{-1}m^{-2}$ and $\mu_{nuc}\sim 1.67\times 10^{-27}kg$ is
nucleon mass. Notice that also in this case no explicit dependence
on mass appears. It turns out that in the ``worst'' NNG case
($\omega _{m}=\omega _{m}^{\exp }\times 10^{-7}$), in which
visibility goes practically to zero, GRW predicts a visibility
lowering at the end of the cycle of about $0.5\%$ with respect to
the free case. It is interesting to note that for smaller $\Delta
x$ an enhancement of the effect's observability comes from the
consideration of granularity. As discussed in the previous
Section, in fact, for $\Delta x\sim 10^{-12}$ or less matter
granularity may come into play; as a consequence, we obtain that
already for  $\omega _{m}=\omega _{m}^{\exp }\times 10^{-3}$
visibility behaves in a similar way to the homogeneous case with
$\omega _{m}=\omega _{m}^{\exp }\times 10^{-6}$ (dashed curve of
Figure \ref{fig:Visibility}).
\bigskip
\section{Monitoring the mirror's state: the Wigner function}

\setcounter{equation}{0} In the previous Section we have
considered a measure of photons' interference, and shown that with
a proper choice of parameters one gets a reduced revival
effect at the end of one oscillation period. Here we want to
elucidate about the mirror's state soon after a photon measurement
process. For monitoring mirror's state, and to get a physical
insight of what's going on, we use the Wigner function. As it is well
known, this quasi-distribution have both positive and negative
parts, the latter being a signature of quantum coherence survival.
It is expected that the action of NNG-induced decoherence would
reduce, after some time, the interference patterns in the Wigner
distribution.

We calculate the Wigner function starting from the expression:
\begin{equation}
W\left( x,p;t\right) =\frac{1}{2\hbar \pi ^{2}}\int d^{2}\lambda \
e^{-\lambda \eta ^{\ast }+\lambda ^{\ast }\eta }\ Tr\left[ \rho _{m}\left(
t\right) \ e^{\lambda \widehat{b}^{\dagger }-\lambda ^{\ast }\widehat{b}}%
\right],  \label{wigner}
\end{equation}
with
\begin{equation*}
\eta =\frac{ip}{\sqrt{2M\omega _{m}^{\ast }\hbar }}+x\sqrt{\frac{M\omega
_{m}^{\ast }}{2\hbar }},
\end{equation*}
where $\rho _{m}$ is the reduced density matrix of the mirror
after a photon detection \cite{30}. Following \cite{31}, we take
this measurement as the process projecting (physical) photons'
state onto the state
\begin{equation*}
|\varphi \rangle =\frac{1}{\sqrt{2}}\left( |0_{A},1_{B}\rangle +e^{i\theta
}|1_{A},0_{B}\rangle \right) ,
\end{equation*}
where $\theta $ is a phase constant (it can be shown that the
corresponding results are quasi-independent of $\theta $ and then
we put $\theta =0$). Correspondingly, mirror's density matrix is
($\left\vert \left\vert \varphi \right\rangle \right\rangle
\equiv\left\vert \varphi \right\rangle \otimes \left\vert
\widetilde{\varphi }\right\rangle $)
\begin{equation*}
\rho _{m}\left( t\right) =Tr_{Ph,\widetilde{Ph},\widetilde{m}}\left[ \left(
\left\vert \left\vert \varphi \right\rangle \right\rangle \left\langle
\left\langle \varphi \right\vert \right\vert \otimes {\large 1}\right)
~\left\vert \left\vert \Psi \left( t\right) \right\rangle \right\rangle
\left\langle \left\langle \Psi \left( t\right) \right\vert \right\vert %
\right] .
\end{equation*}

After some calculations, we get the Wigner function as

\begin{widetext}
\begin{eqnarray}
W\left( x,p;t\right) &=&\frac{1}{2\hbar \pi ^{2}}\int d^{2}\lambda \ \left[
e^{-\lambda \eta ^{\ast }+\lambda ^{\ast }\eta }\ Tr\left[ \rho _{m}\left(
t\right) \ e^{\lambda b^{\dagger }-\lambda ^{\ast }b}\right] \right] =
\notag \\
&=&\frac{1}{4\hbar \pi ^{5}}\int d^{2}(\beta ,\widetilde{\beta },\beta
^{\prime },\widetilde{\beta ^{\prime }})e^{-2\eta \eta ^{\ast }}\biggl[%
\alpha _{1}e^{-\beta _{c}\beta _{c}^{\prime ^{\ast }}-\frac{1}{2}\left\vert
\beta _{c}\right\vert ^{2}-\frac{1}{2}\left\vert \beta _{c}^{\prime
}\right\vert ^{2}+2\beta _{c}\eta ^{\ast }+2\beta _{c}^{\prime ^{\ast }}\eta
}+\alpha _{2}e^{-\beta _{l}\beta _{l}^{\prime ^{\ast }}-\frac{1}{2}%
\left\vert \beta _{l}\right\vert ^{2}-\frac{1}{2}\left\vert \beta
_{l}^{\prime }\right\vert ^{2}+2\beta _{l}\eta ^{\ast }+2\beta _{l}^{\prime
^{\ast }}\eta }  \notag \\
&+&\alpha _{3}e^{-\beta _{c}\beta _{l}^{\prime ^{\ast }}-\frac{1}{2}%
\left\vert \beta _{c}\right\vert ^{2}-\frac{1}{2}\left\vert \beta
_{l}^{\prime }\right\vert ^{2}+2\beta _{c}\eta ^{\ast }+2\beta _{l}^{\prime
^{\ast }}\eta }+\alpha _{4}e^{-\beta _{l}\beta _{c}^{\prime ^{\ast }}-\frac{1%
}{2}\left\vert \beta _{l}\right\vert ^{2}-\frac{1}{2}\left\vert \beta
_{c}^{\prime }\right\vert ^{2}+2\beta _{l}\eta ^{\ast }+2\beta _{c}^{\prime
^{\ast }}\eta }\biggr].  \label{Wigner_int}
\end{eqnarray}
\end{widetext}
Explicit expressions for $\alpha _{1},\alpha _{2},\alpha _{3}$ and
$\alpha _{4}$, together with calculational details, are reported
in Appendix D.

Results for $\kappa =2$ and $\omega _{m}=\omega _{m}^{\exp }\times
10^{-5}$ and\ $\omega _{m}=~5\,\omega _{m}^{\exp }\times 10^{-7}$,
are shown in Figures \ref{fig:Wig_func_1} and \ref{fig:Wig_func_2}
respectively for the homogeneous case. As seen for visibility, in
the case of (proposed) experimental values, Wigner function after
measurement is undistinguishable from the free case. For the
second (presently unattainable) much smaller value of $\omega
_{m}$, after a certain time a diminution of interference fringes
is observed together with a contextual lowering of the first rest
peak.

\begin{widetext}
\begin{figure}[htbp]
\centering \subfigure[] {\includegraphics[height=35mm]{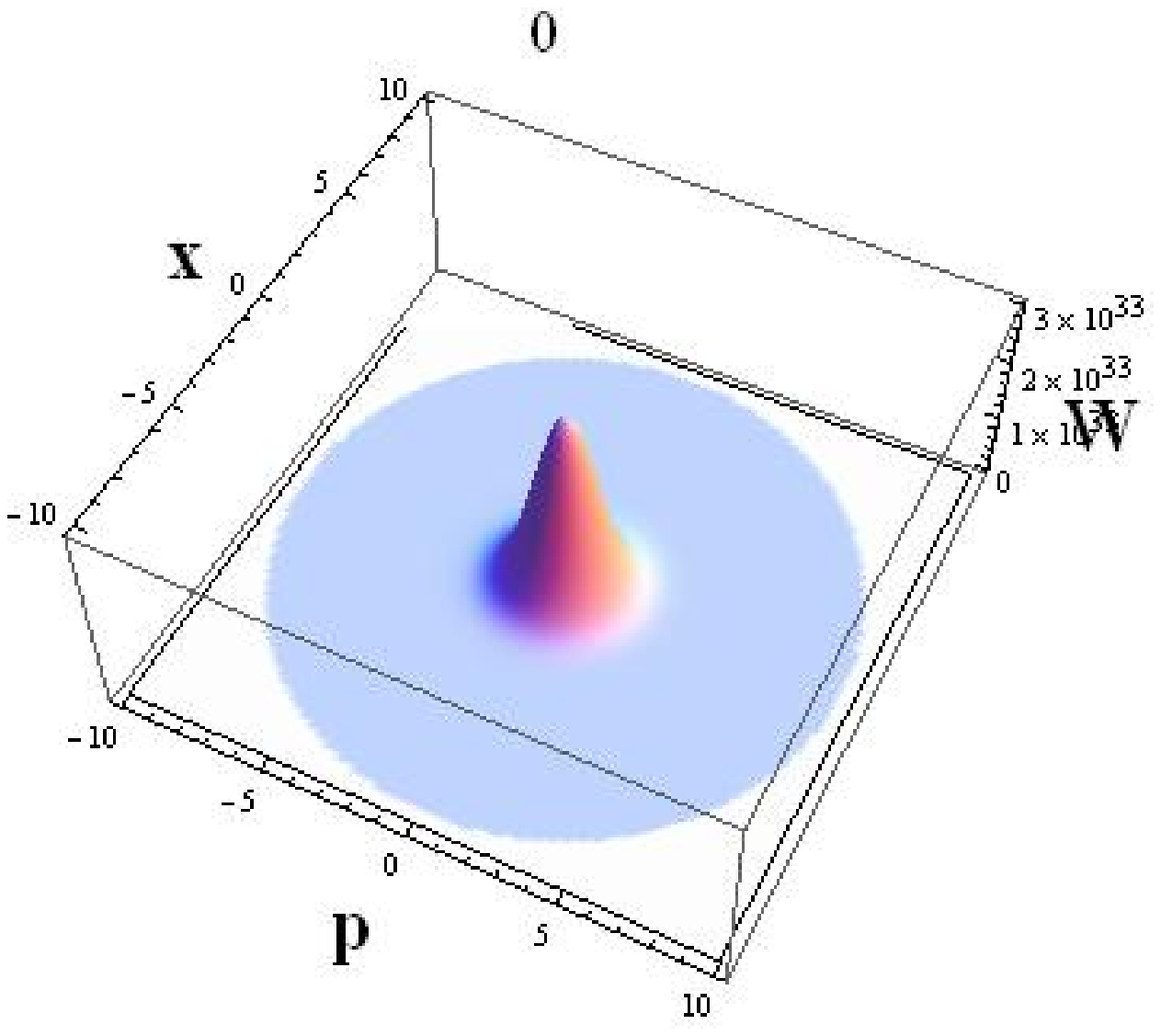}}
\qquad\qquad\subfigure[]
{\includegraphics[height=35mm]{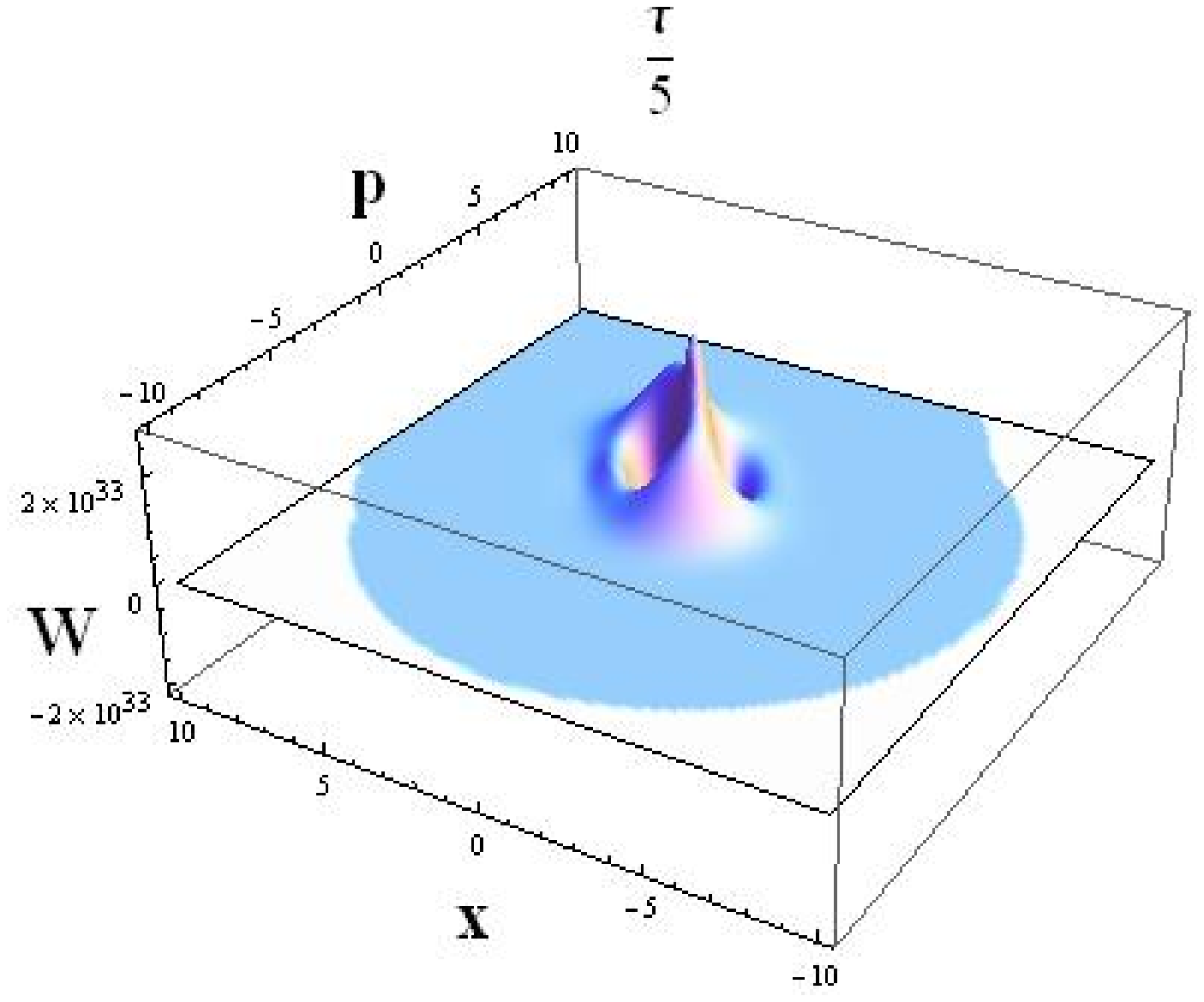}}\qquad\qquad\subfigure[]
{\includegraphics[height=40mm]{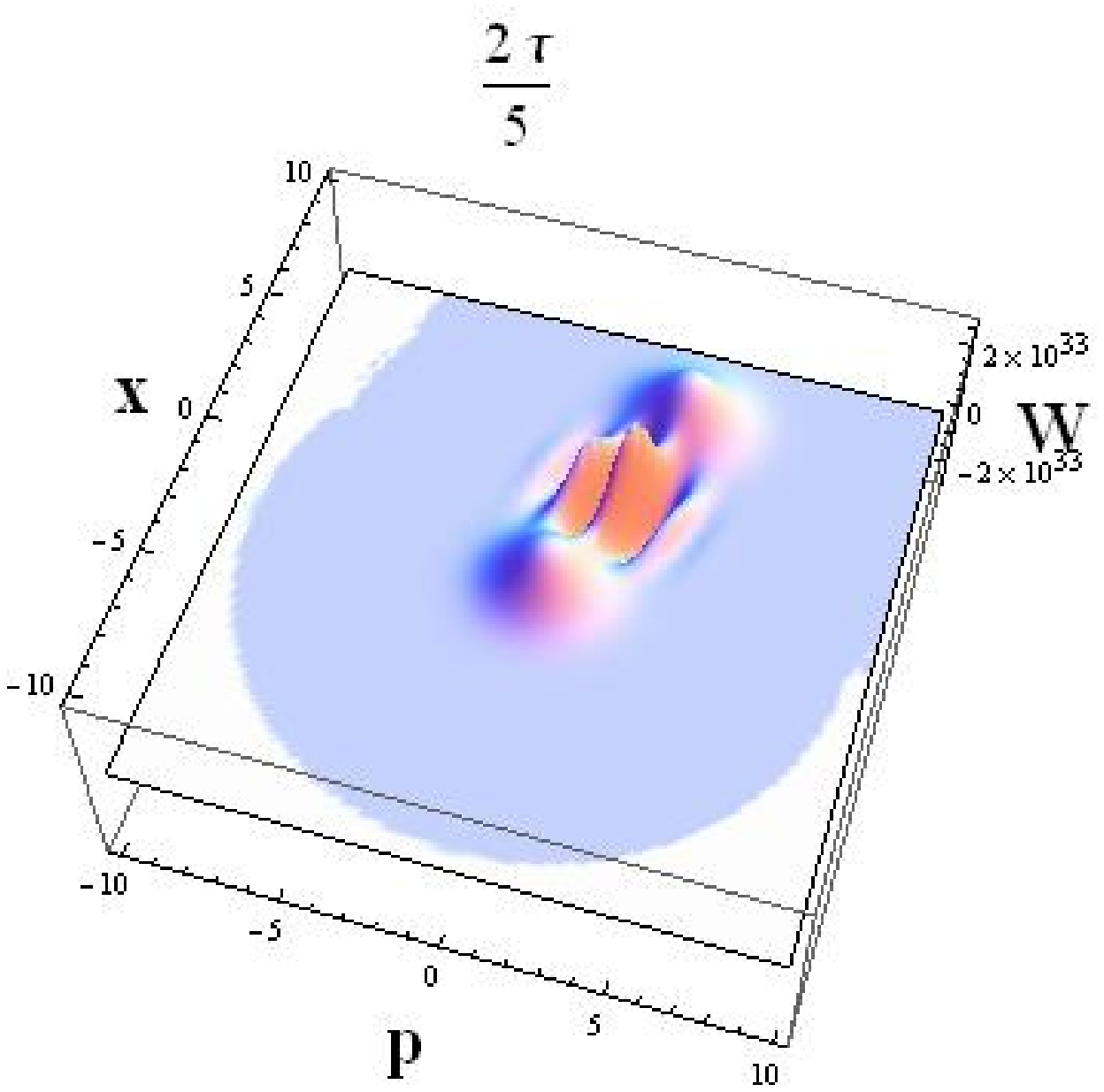}}\qquad\qquad\subfigure[]
{\includegraphics[height=40mm]{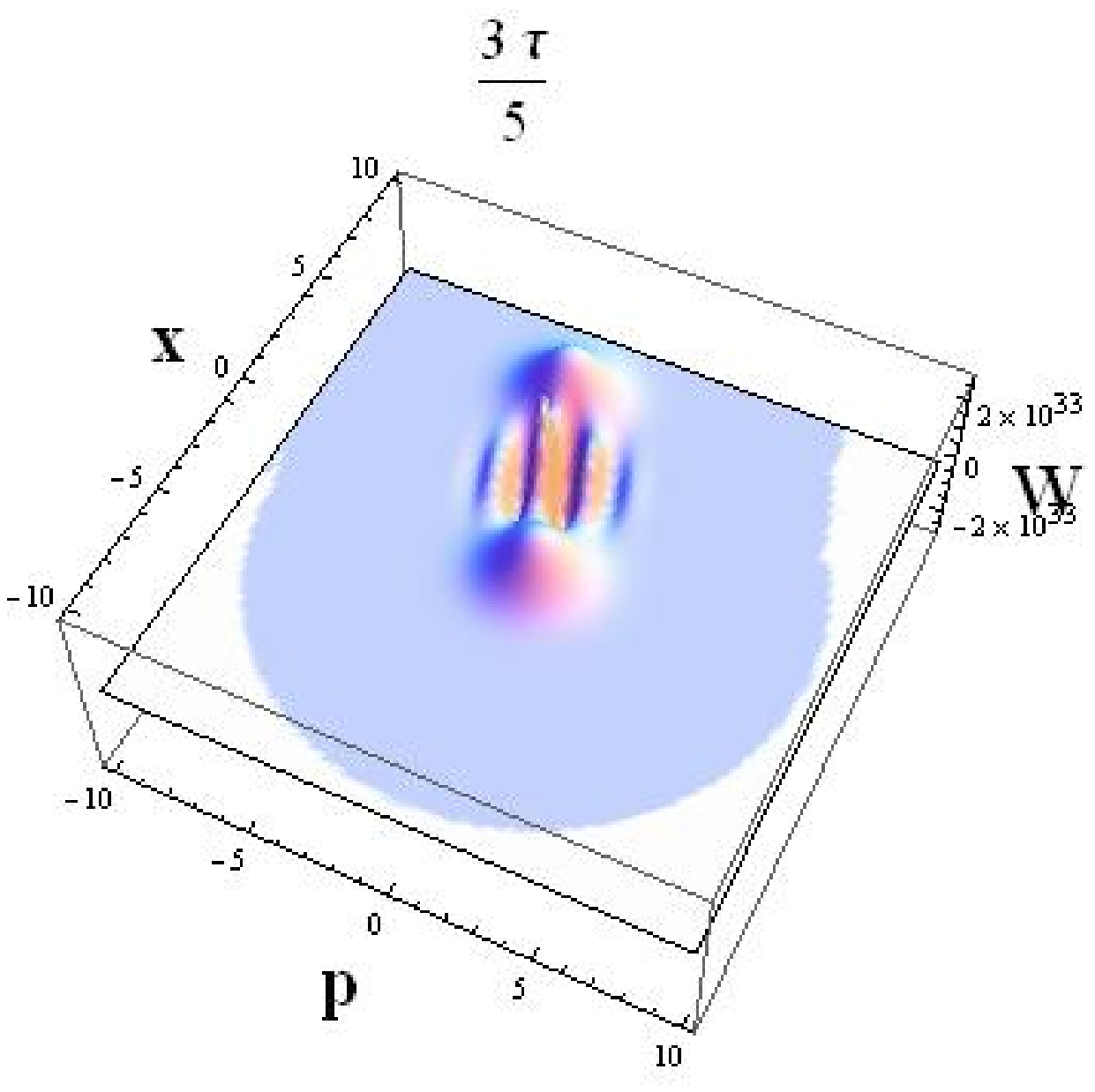}}\qquad\qquad\subfigure[]
{\includegraphics[height=40mm]{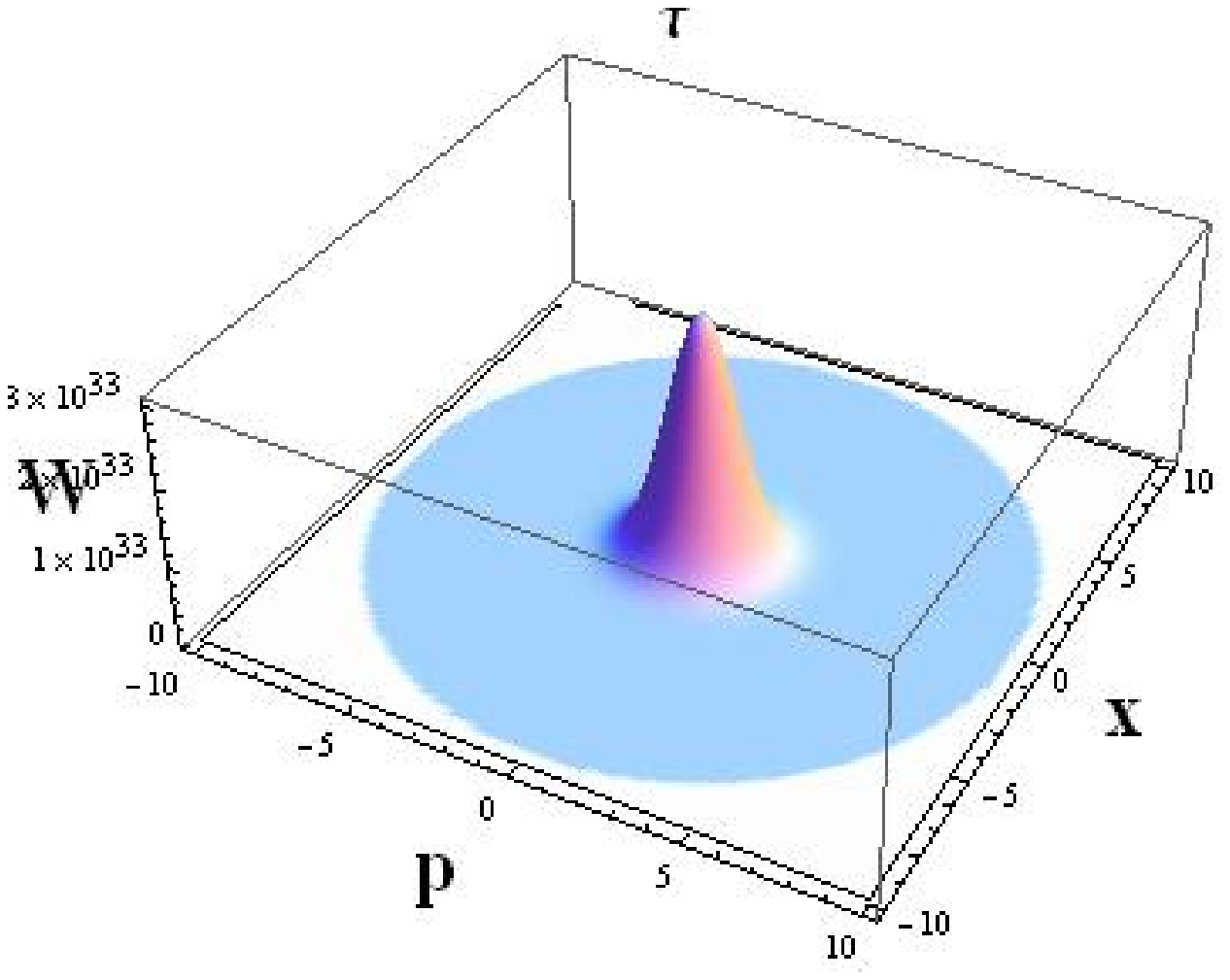}} \caption{(Color online) Wigner function
in the homogeneous case for $\kappa=~2,$
$\omega_m=\omega_m^{exp}\times 10^{-5},$ for the mirror size
$L=10^{-5}m$ and for different intermediate times in a complete
mirror oscillation. The variables $x$ and $p$ are in the ranges
$\{-10 \delta x, 10 \delta x\}$ and $\{-10 \delta p,10 \delta p\}$
with $\delta x\doteq\sqrt{\frac{\hbar}{2 M \omega_m^*}}$ and
$\delta p\doteq\sqrt{\frac{\hbar M \omega_m^*}{2}}$, while $\tau=\omega_m^* t$. In the case
with $\omega_m=\omega_m^{exp},$ as in the absence of gravity, we
get qualitatively the same graphics. All quantities are expressed in dimensionless units.} \label{fig:Wig_func_1}
\end{figure}
\begin{figure}[htbp]
\centering\subfigure[]{\includegraphics[height=35mm]{0.eps}}\qquad\qquad
\subfigure[]
{\includegraphics[height=40mm]{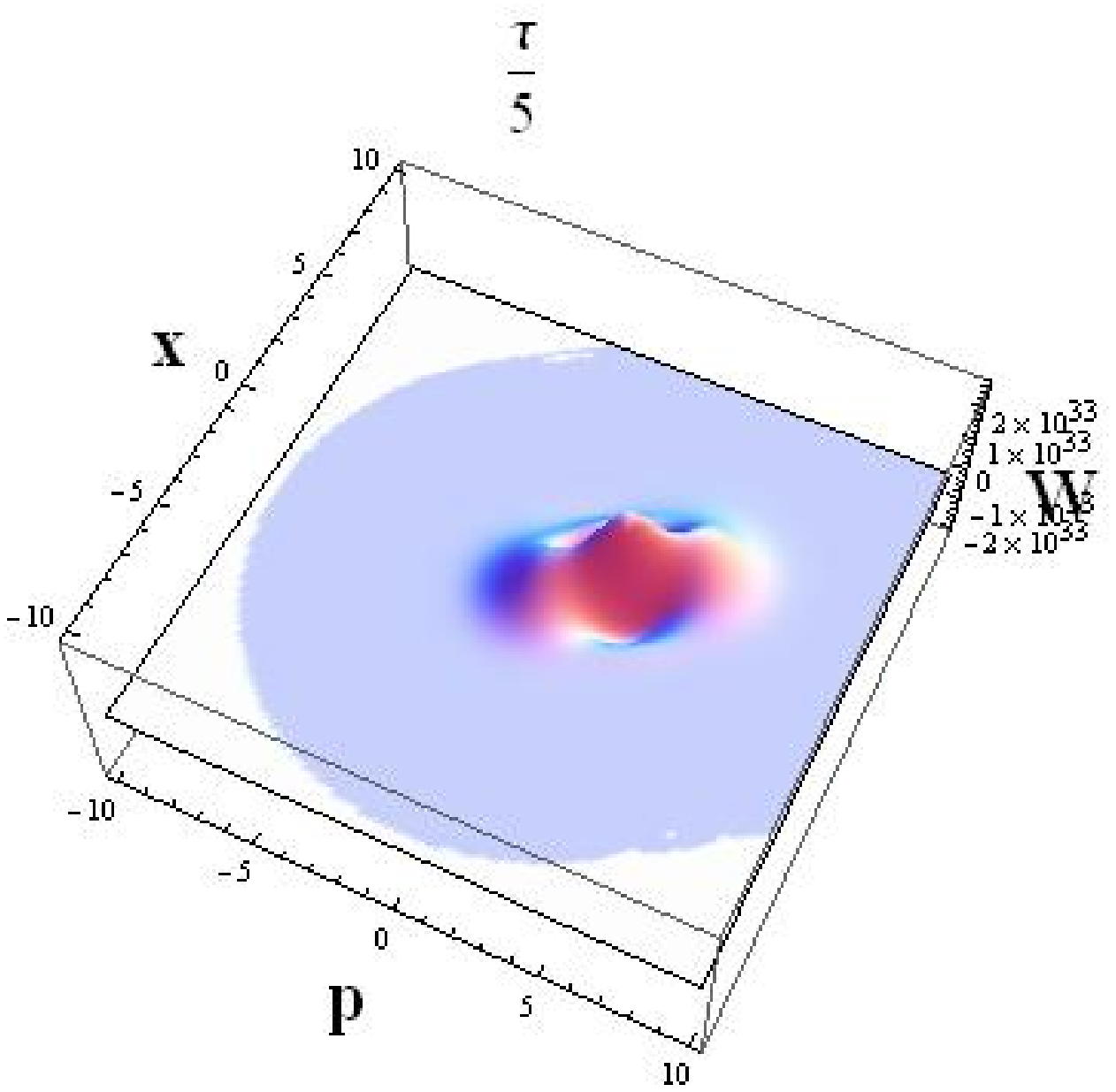}}\qquad\qquad\subfigure[]
{\includegraphics[height=40mm]{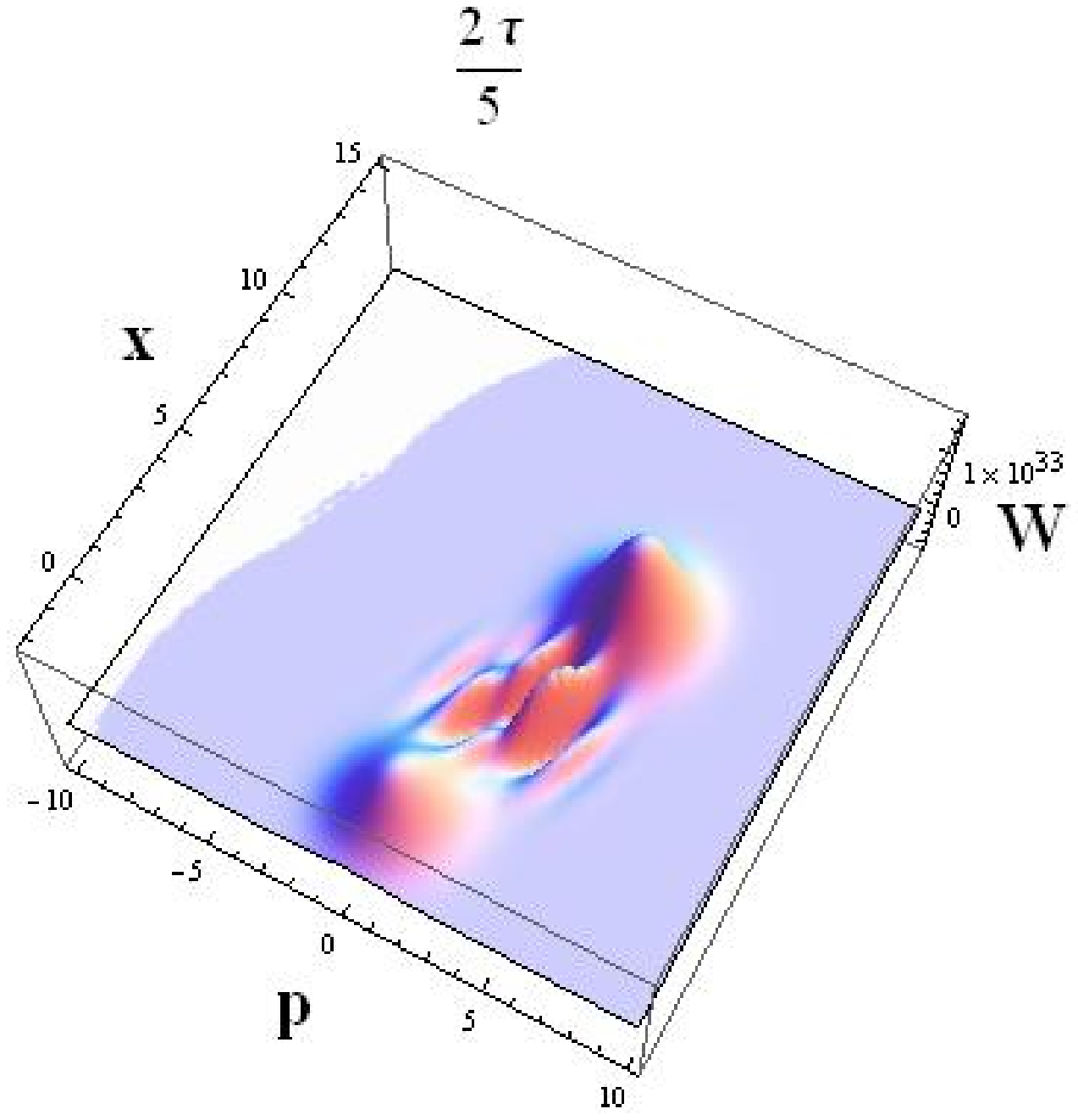}}\qquad\qquad\subfigure[]
{\includegraphics[height=40mm]{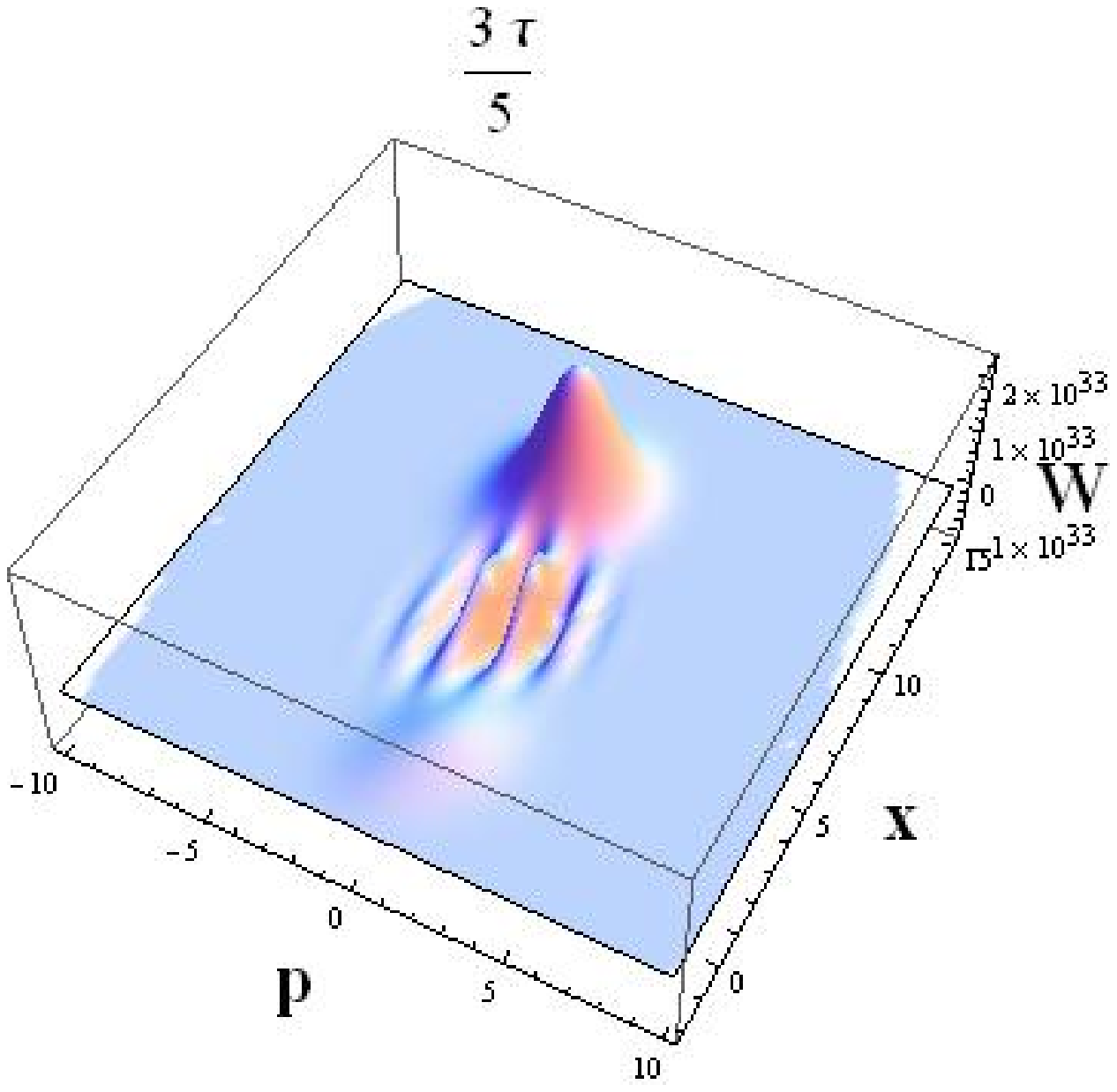}}\qquad\qquad\subfigure[]
{\includegraphics[height=40mm]{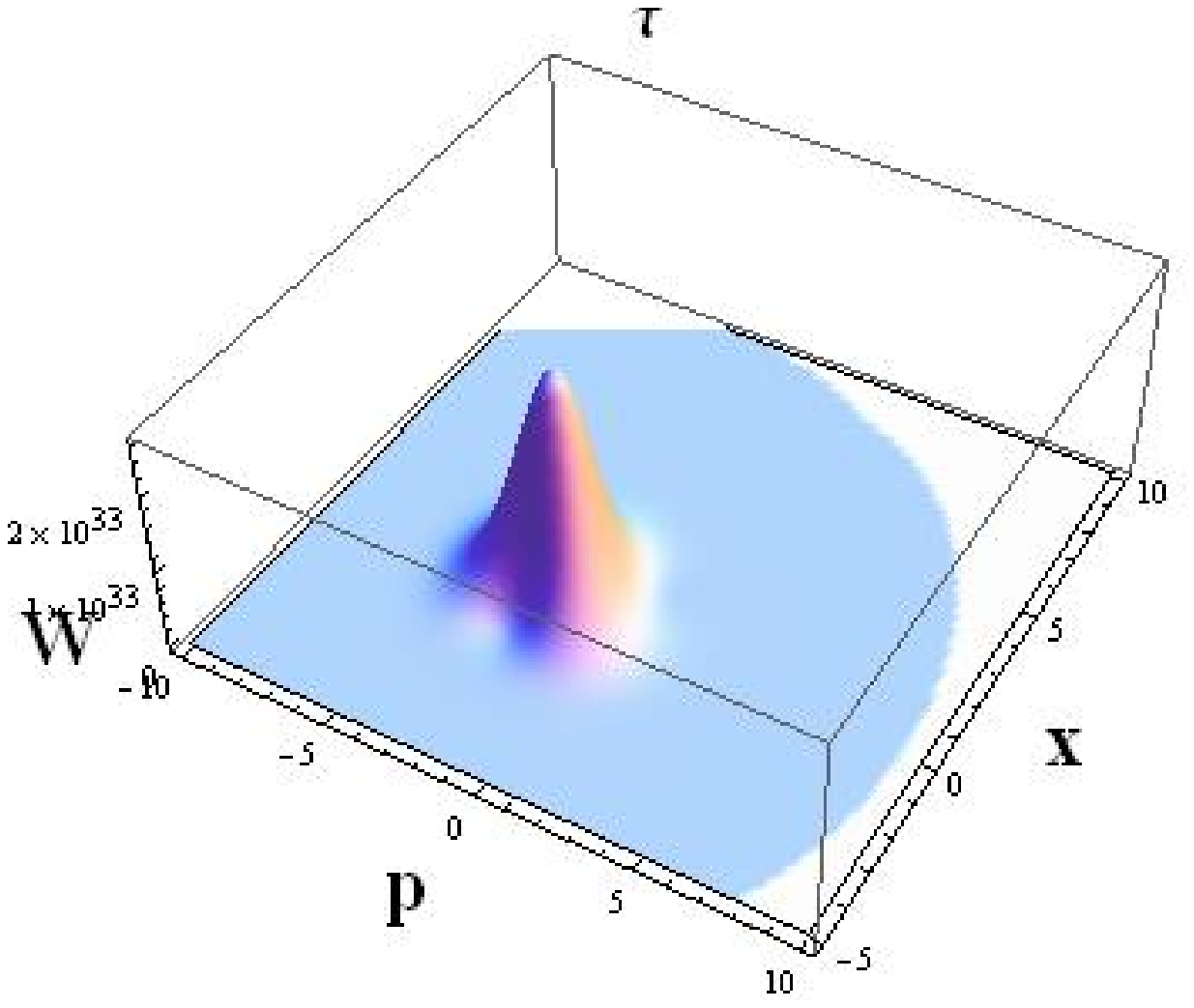}} \caption{(Color online) Wigner function
for $\omega_m=~5\, \omega_m^{exp}\times 10^{-7}$. All the other
parameters are as above. All quantities are expressed in dimensionless units.} \label{fig:Wig_func_2}
\end{figure}
\end{widetext}

\section{Conclusions and perspectives}
\setcounter{equation}{0} In this paper the output of the mirror
experiment proposed by Marshall et al. \cite{29} has been
calculated within the framework of NNG, assuming both homogeneous
and granular mass distributions. By varying the experimental
parameters in a wide range beyond the proposed values \cite{29}, a
window of \textquotedblleft sensible\textquotedblright\ parameters
has been found in which the NNG induced decoherence effect is
manifest.

In conclusion, even if the experimental test of nonunitary gravity
has been proved to lie beyond current technology yet, requiring an
unprecedented control of decoherence, its peculiar form of
self-gravitational interaction has been shown to be in principle
distinguishable from the action of other collapse models. An
exploration of the relevant parameter space could in fact, in a
feasible experiment, lead to a clear distinction of the most
appropriate model. The signature of NNG model is ultimately
connected with the fact that fundamental interaction occurs with a
\textquotedblleft simple\textquotedblright\ system (the mirror's
copy in this case) rather than with a fundamental
\textquotedblleft thermal bath\textquotedblright\ random field
leading to a visibility output somehow indistinguishable from the
effect of temperature. This is essentially due to the laboratory
artificially created superposition state, while in naturally
occurring circumstances it is expected that \textquotedblleft
fundamental environment\textquotedblright , being as complex as
the system itself, could eventually lead to auto-thermalization
effects.\\
In spite of the huge technical challenges, however, we believe
that due to the rapid progress in developing high-quality
micro-optomechanical devices, a prototypal experiment of this type
could be soon realized.

A remark is finally in order concerning a finite temperature
inclusion into the model. It should be clear that, when our initial
knowledge of the system state is characterized by a density matrix
like a thermal state, there is no unique prescription to associate
it with a pure meta-state. In such a case one has to consider the
possibility of using mixed meta-states to encode our incomplete
knowledge. This more general case, independently of the specific
experiment treated here, will pave the way towards a generalized model
of gravity induced thermalization. Such an interesting issue will be addressed in a future publication \cite{noi}.
\section*{Acknowledgements}
The authors would like to thank Mario Salerno for discussions and
useful comments on the manuscript. Anonymous referee's critical
comment, pointing out an important issue of the work, is also
kindly acknowledged.
\bigskip
\appendix
\section{Interacting gravitational potential of the meta-mirrors}

\setcounter{equation}{0} In this Appendix the gravitational
interaction potential between meta-mirrors is computed within the
two assumption of homogeneous and granular mass distributions.
\subsection{The case with homogeneous masses}
Let's consider the Newtonian potential energy for two particles
with masses $M_{1}$ and $M_{2}$
\begin{equation*}
V\left( r_{12}\right) =-G\frac{M_{1}M_{2}}{r_{12}}.
\end{equation*}

Starting from the initial condition where the two mirrors of
length $L$ are overlapped, we consider the shifting of the first
mirror of $d/2$ along $z_{1}$ positive axis and the second one
along $z_{2}$ negative axis. Because of the very small relative
displacement $d$ between the meta-mirrors (which is at maximum of
the order of the size of the wave-function describing, in the
ordinary setting, the CM coordinate of the mirror), it is
enough to calculate the quadratic term of the expansion in the
distance of the total gravitational interaction energy
\begin{widetext}
\begin{eqnarray*}
V\left( d\right) &=&\int_{-\left( L+d\right) /2}^{\left(
L-d\right) /2}dz_{1}\int_{-(L-d)/2}^{\left( L+d\right)
/2}dz_{2}\;\mathcal{V}\left(
z_{1},z_{2}\right) \\
&=&\left(
\int_{-L/2}^{L/2}dz_{1}+\int_{-(L+d)/2}^{-L/2}dz_{1}-\int_{\left(
L-d\right) /2}^{L/2}dz_{1}\right) \left(
\int_{-L/2}^{L/2}dz_{2}-\int_{-L/2}^{-\left( L-d\right)
/2}dz_{2}+\int_{L/2}^{\left( L+d\right) /2}dz_{2}\right)
\mathcal{V}\left(
z_{1},z_{2}\right) \\
&=&\mbox{\textit{Const.}}+\frac{1}{2}\left[ \mathcal{V}\left( \frac{L}{2},-\frac{L}{2}%
\right) -\mathcal{V}\left( \frac{L}{2},\frac{L}{2}\right) \right]
d^{2}+O\left( d^{3}\right),
\end{eqnarray*}
\end{widetext}
where $\mathcal{V}\left( z_{1},z_{2}\right) \delta z_{1}\delta
z_{2}$ is the interaction energy of two square infinitesimally
tiny mirror sheets parallel to the $x-y$ plane,
$$\mathcal{V}\left( z_{1},z_{2}\right)
=\iint\limits_{-\frac{L}{2}}^{\frac{L}{2}}d\,x_{1}d\,y_{1}\iint\limits_{-%
\frac{L}{2}}^{\frac{L}{2}}d\,x_{2}d\,y_{2}\left( -\frac{G\,\rho ^{2}}{r_{12}}%
\right).$$ Here we have used the assumption that, for example, $
\int_{-(L+d)/2}^{-L/2}dz_{1}\int_{-L/2}^{L/2}dz_{2}\mathcal{V}(z_{1},z_{2})=~
\frac{d}{2}\int_{-L/2}^{L/2}\mathcal{V}(-\frac{L}{2},z_{2})dz_2.$
In this way the terms linear in $d$ vanish. Defining the
non-dimensional coordinates $x_{1}^{\prime }=x_{1}/L,\
y_{1}^{\prime }=y_{1}/L,\ $and so on, we get:
\begin{widetext}
\begin{eqnarray*}& &\left[ \mathcal{V}\left( \frac{L}{2},-\frac{L}{2}\right)
-\mathcal{V}\left( \frac{L}{2},\frac{L}{2}\right)
\right]=\\
&=&\frac{G\,M^{2}}{L^{3}} \iint\limits_{-1/2}^{1/2}dx_{1}^{\prime
}dy_{1}^{\prime }\iint\limits_{-1/2}^{1/2}dx_{2}^{\prime
}dy_{2}^{\prime }\left( \frac{1}{ \left[ \left( x_{1}^{\prime
}-x_{2}^{\prime }\right) ^{2}+\left( y_{1}^{\prime }-y_{2}^{\prime
}\right) ^{2}\right] ^{1/2}}-\frac{1}{\left[ \left( x_{1}^{\prime
}-x_{2}^{\prime }\right) ^{2}+\left( y_{1}^{\prime }-y_{2}^{\prime
}\right) ^{2}+1\right] ^{1/2}}\right)=\\
&=&4\frac{G\,M^{2}}{L^{3}}\int_{0}^{1}d\xi _{-}\int_{0}^{1-\xi
_{-}}d\xi _{+}\int_{0}^{1}d\eta _{-}\int_{0}^{1-\eta _{-}}d\eta
_{+}\left( \frac{1}{ \left[ \xi _{-}^{2}+\eta _{-}^{2}\right]
^{1/2}}-\frac{1}{\left[ \xi
_{-}^{2}+\eta _{-}^{2}+1\right] ^{1/2}}\right) \\
&=&4\frac{G\,M^{2}}{L^{3}}\int_{0}^{1}d\xi _{-}\int_{0}^{1}d\eta
_{-}\left( 1-\xi _{-}\right)
\left( 1-\eta _{-}\right) \left( \frac{1}{\left[ \xi _{-}^{2}+\eta _{-}^{2}%
\right] ^{1/2}}-\frac{1}{\left[ \xi _{-}^{2}+\eta _{-}^{2}+1\right] ^{1/2}}%
\right) =\frac{2\pi }{3}\frac{G\,M^{2}}{L^{3}},
\end{eqnarray*}
\end{widetext}
where we have introduced the new variables $\xi _{\pm
}=x_{1}^{\prime }\pm x_{2}^{\prime }$ and $\eta _{\pm
}=y_{1}^{\prime }\pm y_{2}^{\prime }$.

Then we obtain

\begin{equation*}
V\left( d\right) =Const.+\frac{\pi }{3}GM\,\rho_{sil}\;d^{2},
\end{equation*}
from which, writing the interaction term in the meta-Hamiltonian
as $ -K_{G}\left( b+b^{\dagger }\right)
(\widetilde{b}+\widetilde{b}^{\dagger })$, we get $$K_{G}^{\hom
}=\frac{\pi \hbar G\rho _{sil}}{3\omega _{m}}.$$
\subsection{The case with matter-granularity effect}
Looking at the meta-mirrors as aggregates of atoms disposed in a
lattice $\{\textbf{R}_j\}$, we are led to consider the
gravitational potential between the meta-crystals described by the
state (to be symmetrized)
\begin{equation*}
\left\vert \mathbf{\Phi }_{Crystal}^{\left( -d/2\right)
}\right\rangle \otimes \left\vert \widetilde{\mathbf{\Phi
}}_{Crystal}^{\left( +d/2\right) }\right\rangle ,
\end{equation*}
where superscripts parameters $\mp d/2$ refer to the center of
mass displacements from the origin of coordinate system along,
say, $x-$axis. The problem is similar to that of potential between
two atoms or molecules, where (within the Born-Oppenheimer
approximation) nuclei positions are treated as parameters of the
atoms/molecules.\\
The interaction potential is then given by ($\mathbf{d}\equiv
\left(d,0,0\right)$)
\begin{widetext}
\begin{eqnarray}
\label{potential_granular}
\begin{split}
V\left( d\right)  &=-Gm_{nuc}^{2}\int \int
d\mathbf{x}d\mathbf{y}\frac{ \left\langle \mathbf{\Phi
}_{Crystal}^{\left( -d/2\right) }\right\vert \psi ^{\dagger
}\left( \mathbf{x}\right) \psi \left( \mathbf{x}\right) \left\vert
\mathbf{\Phi }_{Crystal}^{\left( -d/2\right) }\right\rangle
\left\langle \widetilde{\mathbf{\Phi }}_{Crystal}^{\left(
+d/2\right) }\right\vert \widetilde{\psi }^{\dagger }\left(
\mathbf{y}\right) \widetilde{\psi }\left( \mathbf{y}\right)
\left\vert \widetilde{\mathbf{\Phi }}_{Crystal}^{\left(
+d/2\right) }\right\rangle }{\left\vert
\mathbf{x}-\mathbf{y}\right\vert } \\
&=-Gm_{nuc}^{2}\left( \frac{m_{nuc}\omega _{Crystal}}{\hbar \pi
}\right) \sum_{h,k=1}^{N_{nuc}}\int \int
d\mathbf{x}d\mathbf{y}\frac{e^{-\frac{ m_{nuc}\omega
_{Crystal}}{\hbar }\left[ \left( \mathbf{x}-\mathbf{R}
_{h}^{(-d/2)}\right) ^{2}+\left(
\mathbf{y}-\mathbf{R}_{k}^{(+d/2)}\right) ^{2}\right]}}{\left\vert
\mathbf{x}-\mathbf{y}\right\vert } \\
&\simeq -Gm_{nuc}^{2}\left( \frac{m_{nuc}\omega _{Crystal}}{\hbar
\pi } \right) N_{nuc}\int \int
d\mathbf{x}d\mathbf{y}\frac{e^{-\frac{
m_{nuc}\omega _{Crystal}}{\hbar }\left[ \left( \mathbf{x}+\mathbf{d}%
/2\right) ^{2}+\left( \mathbf{y}-\mathbf{d}/2\right)
^{2}\right]}}{ \left\vert \mathbf{x-y}\right\vert },
\end{split}
\end{eqnarray}
\end{widetext}
where, for simplicity, Einstein model for the crystal has been
used, with $\omega _{Crystal}\equiv$ Einstein frequency $\simeq
10\,THz$, $m_{nuc}\simeq 4.7\times 10^{-26}Kg$ is nucleus mass,
$N_{nuc}\simeq 10^{14}$ the number of nuclei. In the last line of
the above formula, we have made the assumption that the nuclei
wavefunction spreads are much lower than interatomic distance, and
that the formers are greater than $d$. To get a simple estimate, we
can consider the interaction between interpenetrating spheres of
radius $\sqrt{\frac{\hbar }{2m_{nuc}\omega _{Crystal}}}$ separated
by a distance $d$, so that
\begin{equation*}
V\left( d\right) \simeq Const.+\frac{2}{3}GM\rho _{nuc}\;d^{2},
\end{equation*}
giving the (enhanced) gravitational coupling
$$K_{G}^{\text{ gran}}=\frac{2\hbar G\rho _{nuc}}{3\omega
_{m}}\simeq 10^{4}\times K_{G}^{\text{hom}}.$$
\section{Dynamical evolution of the meta-system}
\setcounter{equation}{0} In the following we study in detail the dynamical evolution of the meta-system.
As initial meta-state we take
\begin{widetext}
\begin{equation}
\left\vert \left\vert \Psi \left( 0\right) \right\rangle
\right\rangle = \frac{1}{2}\left[ \left( \left\vert
0_{A}1_{B}\right\rangle +\left\vert 1_{A}0_{B}\right\rangle
\right) \ \left\vert \alpha\right\rangle \right] \otimes \left[
\left( \left\vert 0_{\tilde{A}}1_{\tilde{B}}\right\rangle
+\left\vert 1_{\tilde{A}}0_{\tilde{B}}\right\rangle \right) \
\left\vert \widetilde{\alpha}\right\rangle
\right]\label{state}\equiv \left\vert \psi \left( 0\right)
\right\rangle \otimes \left\vert \psi \left( 0\right)
\right\rangle
\end{equation}
\end{widetext}where general coherent states $\alpha _{1}$ and $\alpha _{2}$
are considered, although acceptable meta-states must be symmetrized with
respect to the physical and hidden parts.

Let's start by calculating the meta-state at time $t$. We will use the
interaction picture, defining
\begin{widetext}
$$\left\vert \left\vert \Psi \left( t\right) \right\rangle
\right\rangle _{Int}=e^{i\left(
H_{free}\left[b,b^{\dag},\mathcal{N}_{A},\mathcal{N} _{B};\omega
_{m}^{\ast }\right] +H_{free}\left[
\tilde{b},\tilde{b}^{\dag},\mathcal{N}_{A}^{\sim
},\mathcal{N}_{B}^{\sim };\omega _{m}^{\ast }\right] \right)
t/\hbar }\left\vert \left\vert \Psi(t)\right\rangle \right\rangle
_{Sch}\,$$ \end{widetext}
where
\begin{equation*}
\left\vert \left\vert \Psi \left( t\right) \right\rangle
\right\rangle _{Sch} =\left\vert \psi \left( t\right)
\right\rangle \otimes \left\vert \psi \left( t\right)
\right\rangle
\end{equation*}
and
\begin{widetext}
\begin{equation}\label{psi}
  \left\vert \psi \left( t\right) \right\rangle =\frac{1}{\sqrt{2}}
e^{-i\omega _{Ph}t}\left(\left\vert 0_{A}1_{B}\right\rangle
\otimes \left\vert \alpha e^{-i\omega _{m}^*t}\right\rangle
+e^{i\kappa ^{2}\left( \omega _{m}^*t-\sin \omega
_{m}^*t\right)}e^{i\kappa\Im[\alpha(1-e^{-i\omega_m^*t})]}
\left\vert 1_{A}0_{B}\right\rangle \otimes \left\vert \alpha
e^{-i\omega _{m}^*t}+\kappa \left( 1-e^{-i\omega _{m}^*t}\right)
\right\rangle\right).
\end{equation}
\end{widetext}
The interaction Hamiltonian
\begin{equation*}
H_{G}=-K_{G}\left( b+b^{\dagger}\right) (\widetilde{b}+\widetilde{b}%
^{\dagger })
\end{equation*}
evolves as
\begin{widetext}
\begin{eqnarray*}
H_{G,Int}\left( t\right) &=&-K_{G}e^{i\left( H_{free}\left[
b,b^{\dagger }, \mathcal{N}_{A},\mathcal{N}_{B};\omega _{m}^{\ast
}\right] +H_{free}\left[ \tilde{b},\tilde{b}^{\dagger
},\mathcal{N}_{A}^{\sim },\mathcal{N}_{B}^{\sim };\omega
_{m}^{\ast }\right] \right) t/\hbar }\left( b+b^{\dagger
}\right)\times\\
&\times& (\widetilde{b}+\widetilde{b}^{\dagger
})e^{-i\left(H_{free}\left[b,b^{\dagger
},\mathcal{N}_{A},\mathcal{N} _{B};\omega _{m}^{\ast }\right]
+H_{free}\left[
\tilde{b},\tilde{b}^{\dagger},\mathcal{N}_{A}^{\sim
},\mathcal{N}_{B}^{\sim };\omega _{m}^{\ast }\right]\right)
t/\hbar}=\\
&=&-K_{G}e^{iH_{free}\left[ b,b^{\dagger
},\mathcal{N}_{A},\mathcal{N} _{B};\omega _{m}^{\ast }\right]
t/\hbar }\left( b+b^{\dagger }\right) e^{-iH_{free}\left[
b,b^{\dagger },\mathcal{N}_{A},\mathcal{N}_{B};\omega _{m}^{\ast
}\right] t/\hbar }\times\\
&\times& e^{iH_{free}\left[ \tilde{b},\tilde{b}^{\dagger },
\mathcal{N}_{A}^{\sim },\mathcal{N}_{B}^{\sim };\omega _{m}^{\ast
}\right] t/\hbar }(\widetilde{b}+\widetilde{b}^{\dagger
})e^{-iH_{free}\left[\tilde{b},\tilde{b}^{\dagger
},\mathcal{N}_{A}^{\sim },\mathcal{N}_{B}^{\sim };\omega
_{m}^{\ast }\right] t/\hbar }.
\end{eqnarray*}
\end{widetext}
The above expression is the product of two specular terms, so it
suffices to calculate the first, say. We make use of the
Backer-Hausdorff lemma:
\begin{equation*}
e^{-F}Ge^{F}=\sum\limits_{n=0}^{\infty
}\frac{\left(-1\right)^{n}}{n!} \left[ F,G\right]_{n},
\end{equation*}
with
\begin{equation*}
\left[F,G\right]_{0}=G,\quad\left[F,G\right] _{n}=\left[
F,\left[F,G\right] _{n-1}\right] ,
\end{equation*}
and
\begin{equation*}
F=-i\left[ \omega _{m}^*b^{\dagger }b-g\mathcal{N}_{A}\left(
b+b^{\dagger }\right) \right] t,\qquad G=b+b^{\dagger },
\end{equation*}
($g$ is now defined as $g=\kappa\,\omega _{m}^*$) by which, noting
that
\begin{widetext}
\begin{eqnarray*}
\left[F,G\right]_{0} &=&G=b+b^{\dagger }, \\
\left[F,G\right]_{1} &=&-i\omega _{m}^*t\left[b^{\dagger
}b,b+b^{\dagger}
\right] =i\omega_{m}^*t\left(b-b^{\dagger}\right) , \\
\left[F,G\right]_{2} &=&\left[ F,\left[F,G\right]_{1}\right]
=-(\omega_{m}^*t)^{2}\left[\left(b+b^{\dagger }\right)-2\left(
g/\omega
_{m}^*\right)\mathcal{N}_{A}\right], \\
\left[F,G\right] _{3} &=&-i(\omega _{m}^*t)^{3}\left( b-b^{\dagger
}\right)
=-(\omega _{m}^*t)^{2}\left[ F,G\right] _{1}, \\
\left[ F,G\right]_{4} &=&-(\omega _{m}^*t)^{2}\left[ F,G\right] _{2} \\
&&... \\
&&... \\
\left[F,G\right]_{n_{even}\neq 0}&=&(-1)^{n_{even}/2}\left(
\omega_{m}^*t\right)^{n_{even}}\left[\left(b+b^{\dagger }\right)
-2\left(g/\omega _{m}^*\right) \mathcal{N}_{A}\right]
\\
\left[ F,G\right]_{n_{odd}} &=&\left( i\omega _{m}^*t\right)
^{n_{odd}}\left( b-b^{\dagger }\right) ,
\end{eqnarray*}
we obtain:

\begin{eqnarray*}\widehat{\mathcal{O}}\left(t\right)&
\equiv & e^{iH_{free}\left[a,a^{\dagger
},\mathcal{N}_{A},\mathcal{N}_{B};\omega _{m}^{\ast }\right]
t/\hbar }\left( b+b^{\dagger }\right) e^{-iH_{free}\left[
a,a^{\dagger }, \mathcal{N}_{A},\mathcal{N}_{B};\omega _{m}^{\ast
}\right] t/\hbar }\\
& = & b^{\dagger } (\cos (\omega _{m}^* t )+i \sin (\omega _{m}^*
t ))+b (\cos (\omega _{m}^* t )-i \sin (\omega _{m}^* t ))-\frac{2
g \mathcal{N}_A \cos (\omega _{m}^* t )}{\omega _{m}^* }+\frac{2 g
\mathcal{N}_A}{\omega _{m}^* },
\end{eqnarray*}
and a similar result for $\widehat{\widetilde{\mathcal{O}}}(t):$

\begin{eqnarray*}\widehat{\widetilde{\mathcal{O}}}\left(t\right)&
\equiv & e^{iH_{free}\left[\widetilde{a},\widetilde{a}^{\dagger
},\widetilde{\mathcal{N}}_{A},\widetilde{\mathcal{N}}_{B};\omega
_{m}^{\ast }\right] t/\hbar }\left(
\widetilde{b}+\widetilde{b}^{\dagger}\right) e^{-iH_{free}\left[
\widetilde{a},\widetilde{a}^{\dagger },
\widetilde{\mathcal{N}}_{A},\widetilde{\mathcal{N}}_{B};\omega
_{m}^{\ast }\right] t/\hbar }\\
& = & \widetilde{b}^{\dagger } (\cos ( \omega _{m}^* t)+i \sin
(\omega _{m}^* t ))+\widetilde{b} (\cos (\omega _{m}^* t )-i \sin
(\omega _{m}^* t ))-\frac{2 g \mathcal{\widetilde{N}}_A \cos
(\omega _{m}^*t )}{\omega _{m}^* }+\frac{2 g
\mathcal{\widetilde{N}}_A}{\omega _{m}^* },
\end{eqnarray*}
or
$$\widehat{\mathcal{O}}(t)
=\gamma(t)b^{\dagger}+\gamma^{\ast}(t)b+\Gamma(t),\qquad
\widetilde{\widehat{\mathcal{O}}}(t)=\gamma(t)\widetilde{b}^{\dagger}+\gamma^{\ast}(t)\widetilde{b}+\widetilde{\Gamma
}(t),$$
where 
\begin{equation*}
\gamma(t)=\cos (\omega _{m}^* t )+i \sin (\omega _{m}^* t ),\quad
\Gamma(t)=-2\kappa \mathcal{N}_A(\cos(\omega _{m}^*t)-1),
\quad\widetilde{\Gamma}(t)=-2\kappa
\mathcal{\widetilde{N}}_A(\cos(\omega _{m}^*t)-1).
\end{equation*}
The evolution equation is 
\begin{equation*}
\frac{d\|\Psi(t)\rangle\rangle_{Int}}{dt}=-\frac{i}{\hbar}
H_{G,Int}(t)\|\Psi(t) \rangle\rangle_{Int}=
\frac{i}{\hbar}K_{G}\widehat{ \mathcal{O}}(t)\widehat{
\widetilde{\mathcal{O}}}(t)\|\Psi(t)\rangle \rangle_{Int},
\end{equation*}
i.e., using the Hubbard-Stratonovich transformation \cite{Negele},
\begin{eqnarray*}
&&\|\Psi
(t)\rangle\rangle_{Int}=\widehat{T}e^{\int_{0}^{t}\frac{i}{\hbar}K_{G}
\widehat{ \mathcal{O}}(t^{\prime})
\widehat{\widetilde{\mathcal{O}}} (t^{\prime
})dt^{\prime}}\|\Psi(0)\rangle=\\
&=&\int D[\varphi_{1}(t),\varphi_{2}(t)]e^{-\frac{ic^{2}}{\hbar}
\int_{0}^{t}dt^{\prime}(\varphi_{1}^{2}-\varphi_{2}^{2})}
\widehat{T}e^{ \frac{ic\sqrt{K_{G}}}{\hbar}
\int_{0}^{t}dt^{\prime}(\varphi_{1}+\varphi_{2})
\widehat{\mathcal{O}}(t)}
\widehat{T}e^{\frac{ic\sqrt{K_{G}}}{\hbar} \int_{0}^{t}dt^{\prime
}(\varphi_{1}-\varphi_{2})\widehat{\widetilde{
\mathcal{O}}}(t)}\|\Psi (0)\rangle\rangle \\
&=&\frac{1}{\pi^2}\iint d^{2}\beta\,d^{2}\widetilde{\beta }\int
D[\varphi_{1}(t)
,\varphi_{2}(t)]e^{-ic^{2}/\hbar\int_{0}^{t}dt^{\prime}(\varphi_{1}^{2}-
\varphi _{2}^{2})}\langle\beta |\widehat{T}e^{ic\sqrt{K_{G}}/\hbar
\int_{0}^{t}dt^{\prime}(\varphi_{1}+\varphi_{2})\widehat{
\mathcal{O}}
(t)}|\alpha\rangle\times \\
&\times& \langle\widetilde{\beta}| \widehat{T}
e^{ic\sqrt{K_{G}}/\hbar \int_{0}^{t}dt^{\prime
}(\varphi_{1}-\varphi_{2})\widehat{ \widetilde{
\mathcal{O}}}(t)}|\widetilde{\alpha}\rangle|\beta\rangle|
ent\rangle\otimes| \widetilde{\beta}
\rangle|\widetilde{ent}\rangle=\frac{1}{\pi^2}\iint d^2\beta
d^2\widetilde{\beta}\;K^{\mathcal{N}_{A} \widetilde{
\mathcal{N}}_{A}}(\beta ,\widetilde{\beta }
;\alpha;t)|\beta\rangle| ent\rangle\otimes|\widetilde{\beta}
\rangle| \widetilde{ent}\rangle,
\end{eqnarray*}
\end{widetext}
where $c$ is a constant. Before dealing with the kernel
$K^{\mathcal{N}_{A} \widetilde{ \mathcal{N}}_{A}}$, we calculate
the amplitude with the help of the Baker-Campbell-Hausdorff
formula:
\begin{equation*}
e^{t(\widehat{A}+\widehat{B})}=e^{t\widehat{A}}e^{t\widehat{B}}e^{-\frac{t^2
}{2}[\widehat{A},\widehat{B}]}
e^{\frac{t^3}{6}(2[\widehat{B},[\widehat{A},
\widehat{B}]]+[\widehat{A},[\widehat{A},\widehat{B}]])}\dots,
\end{equation*}
\begin{widetext}
\begin{eqnarray*}
& &\langle\beta |\widehat{T}e^{\frac{ic\sqrt{K_{G}}}{\hbar}
\int_{0}^{t}dt^{\prime}(\varphi_{1}+\varphi_{2})\widehat{
\mathcal{O}}
(t)}|\alpha\rangle= \\
&=& \langle\beta |\widehat{T} \exp \biggr\{f\,b^{\dagger
}\int_{0}^{t}dt^{\prime}(\varphi_{1}+\varphi_{2})\gamma(t)\biggr\}
\exp
\biggl\{f\,b\int_{0}^{t}dt^{\prime}(\varphi_{1}+\varphi_{2})\gamma^{\ast}(t)
\biggr\} \times \\
&\times&
\exp\biggl\{\frac{f^{2}}{2}\biggl(\int_{0}^{t}dt^{\prime}(
\varphi_{1}+\varphi_{2})
\gamma(t)\biggr)\biggl(\int_{0}^{t}dt^{\prime}(
\varphi_{1}+\varphi_{2}) \gamma^{\ast}(t)\biggr)
\biggr\}|\alpha\rangle\exp
\biggl\{f\int_{0}^{t}dt^{\prime}(\varphi_{1}+\varphi_{2})\Gamma(t)\biggl\}=
\\
&=&e^{f\beta^{\ast}\int_{0}^{t}dt^{\prime}(\varphi_{1}+\varphi_{2})
\gamma(t)}e^{f\alpha
\int_{0}^{t}dt^{\prime}(\varphi_{1}+\varphi_{2})\gamma^{\ast}(t)}
e^{\frac{
f^{2}}{2}\bigl(\int_{0}^{t}dt^{\prime}(\varphi_{1}+\varphi_{2})
\gamma(t)
\bigr)\bigl(\int_{0}^{t}dt^{\prime}(\varphi_{1}+\varphi_{2})
\gamma^{\ast}(t)
\bigr)}e^{f\int_{0}^{t}dt^{\prime}(\varphi_{1}+\varphi_{2})
\Gamma(t)}\langle\beta|\alpha\rangle
\end{eqnarray*}
\end{widetext}
\begin{equation*}
\biggl(f\equiv \frac{ic}{\hbar}\sqrt{K_{G}}\biggr).
\end{equation*}
A similar result holds on for the other amplitude by making the
following substitutions: $ (\varphi_{1}+~\varphi_{2})\rightarrow~
(\varphi_{1}-~\varphi_{2}),\quad\beta^{\ast}\rightarrow~
\widetilde{\beta}
^{\ast},\quad\Gamma\rightarrow\widetilde{\Gamma}.$
We get then
\begin{widetext}
\begin{eqnarray*}
& &K^{\mathcal{N}_{A} \widetilde{\mathcal{N}}_{A}}(\beta,\widetilde{\beta};\alpha;t)= \lim_{\substack{ %
N\rightarrow\infty  \\ \Delta t\rightarrow 0}}\iint\dots\int
\prod_{i=1}^Nd\varphi_{2,i}\iint\dots\int\prod_{i=1}^N d \varphi_{1,i} \exp%
\biggl\{-\frac{i\,c^2}{\hbar}\sum_{i=1}^N\Delta
t(\varphi_{1,i}^2-\varphi_{2,i}^2)+ \\
&+&f\sum_{i=1}^N\Delta
t(\varphi_{1,i}+\varphi_{2,i})(\beta^{\ast}\gamma_{i}+\alpha\gamma^{%
\ast}_{i}+\Gamma_i)+\frac{f^2}{2}\sum_{i,j=1}^N(\Delta
t)^2(\varphi_{1,i}+\varphi_{2,i})\gamma_{i}\gamma^{\ast}_{j}(\varphi_{1,j}+%
\varphi_{2,j})\biggr\}\times \\
&\times& \exp\biggl\{f\sum_{i=1}^N\Delta t(\varphi_{1,i}-\varphi_{2,i})(%
\widetilde{\beta}^{\ast}\gamma_{i}+\alpha\gamma^{\ast}_{i}+\widetilde{\Gamma}%
_i) +\frac{f^2}{2}\sum_{i,j=1}^N(\Delta
t)^2(\varphi_{1,i}-\varphi_{2,i})\gamma_{i}\gamma^{\ast}_{j}(\varphi_{1,j}-%
\varphi_{2,j})\biggr\} \langle\beta|\alpha\rangle\langle\widetilde{\beta}|%
\widetilde{\alpha}\rangle= \\
&=&\exp{\left[-\frac{|\beta|^2}{2}-\frac{|\alpha|^2}{2}+\beta^*\alpha -\frac{%
|\widetilde{\beta}|^2}{2}-\frac{|\alpha|^2}{2}+\widetilde{\beta}^*\alpha%
\right]}\times \\
&\times&\lim_{\substack{ N\rightarrow\infty  \\ \Delta t\rightarrow 0}}%
\iint\dots\int \prod_{i=1}^Nd\varphi_{2,i}\exp\biggl\{\sum_{i=1}^N\biggl[%
\frac{ic^2}{\hbar} \Delta t\delta_{i,j}+\frac{f^2}{2}(\Delta
t)^2\gamma_{i}\gamma^{\ast}_{j}\biggr]\varphi_{2,i}\varphi_{2,j}+
f\sum_{i=1}^N\Delta t\bigl(\beta^{\ast}\gamma_{i}+\Gamma_i-\widetilde{\beta}%
^{\ast}\gamma_{i}-\widetilde{\Gamma}_i\bigr)\varphi_{2,i}\biggr\}\times \\
&\times&\iint\dots\int\prod_{i=1}^Nd\varphi_{1,i}\exp\biggl\{\sum_{i,j=1}^N%
\biggl[-\frac{ic^2}{\hbar}\Delta t\delta_{i,j}+\frac{f^2}{2}(\Delta
t)^2\gamma_{i}\gamma^{\ast}_{j}\biggr]\varphi_{1,i}\varphi_{1,j}+f%
\sum_{i=1}^N\Delta
t(\beta^{\ast}\gamma_{i}+2\alpha\gamma^{\ast}_{i}+\Gamma_i+\widetilde{\beta}%
^{\ast}\gamma_{i} +\widetilde{\Gamma}_i)\varphi_{1,i} \biggl\}= \\
&=&e^{-\frac{|\beta|^2}{2}-\frac{|\alpha|^2}{2}+\beta^*\alpha} e^{-\frac{|%
\widetilde{\beta}|^2}{2}-\frac{|\alpha|^2}{2}+\widetilde{\beta}%
^*\alpha}\times\lim_{\substack{ N\rightarrow\infty  \\ \Delta t\rightarrow 0
}}\sqrt{\frac{(2\pi)^{2N}}{\det\mathbf{A}^{(N)}}}\;e^{-\frac{1}{2}\mathbf{J}%
^T(\mathbf{A}^{(N)})^{-1}\mathbf{J}}= \\
&=&e^{-\frac{|\beta|^2}{2}-\frac{|\alpha|^2}{2}+\beta^*\alpha} e^{-\frac{|%
\widetilde{\beta}|^2}{2}-\frac{|\alpha|^2}{2}+\widetilde{\beta}%
^*\alpha}\times\lim_{\substack{ N\rightarrow\infty  \\ \Delta t\rightarrow 0
}}\frac{(2\pi)^N}{\sqrt{\det\mathbf{A}_1^{(N)}\det\mathbf{A}_2^{(N)}}}\; e^{-%
\frac{1}{2}\mathbf{J}_1^T(\mathbf{A}_1^{(N)})^{-1}\mathbf{J}_1-\frac{1}{2}%
\mathbf{J}_2^T(\mathbf{A}_2^{(N)})^{-1}\mathbf{J}_2} \\
&=&\Lambda(t)\;e^{-\frac{|\beta|^2}{2}-\frac{|\alpha|^2}{2}+\beta^*\alpha}
e^{-\frac{|\widetilde{\beta}|^2}{2}-\frac{|\alpha|^2}{2}+\widetilde{\beta}%
^*\alpha}\times \\
&\times&e^{-\frac{K_G}{2\hbar^2}\int_0^t\!\!\int_0^tdt'dt''[\beta^{\ast}\gamma(t')+
2\alpha\gamma^{\ast}(t')+\Gamma(t')+\widetilde{\beta}^{\ast}\gamma(t')+\widetilde{\Gamma}(t')]\textbf{A}_1^{-1}(t',t'')
[\beta^{\ast}\gamma(t'')+2\alpha\gamma^{\ast}(t'')+\Gamma(t'')+\widetilde{\beta}^{\ast}\gamma(t'')+\widetilde{\Gamma}(t'')]}\times\\
&\times&
e^{-\frac{K_G}{2\hbar^2}\int_0^t\!\!\int_0^tdt'dt''[\beta^{\ast}\gamma(t')
+\Gamma(t')-\widetilde{\beta}^{\ast}\gamma(t')-\widetilde{\Gamma}(t')]\textbf{A}_2^{-1}(t',t'')
[\beta^{\ast}\gamma(t'')+\Gamma(t'')-\widetilde{\beta}^{\ast}\gamma(t'')-\widetilde{\Gamma}(t'')]}.
\end{eqnarray*}
where
\begin{equation*}
\varphi _{1,1}=\varphi _{1}(t=0),\quad \varphi _{1,N}=\varphi _{1}(t),\quad
\varphi _{2,1}=\varphi _{2}(t=0),\quad \varphi _{2,N}=\varphi _{2}(t);\qquad
\Delta t=\frac{t}{N},\ \ \tau =\omega _{m}^{\ast }t\ ,
\end{equation*}
and
\begin{equation*}
\mathbf{A}^{\left( N\right) }=
\begin{pmatrix}
\mathbf{A}_{1}^{\left( N\right) } & 0 \\
0 & \mathbf{A}_{2}^{\left( N\right)}%
\end{pmatrix}%
, \qquad\mathbf{J}^{\left( N\right) }=
\begin{pmatrix}
\mathbf{J}_{1}^{\left( N\right) } \\
\mathbf{J}_{2}^{\left( N\right) }%
\end{pmatrix}%
,
\end{equation*}
\begin{equation*}
\left[ \mathbf{A}_{1}^{\left( N\right) }\right] _{ij}=\omega _{m}^{\ast
-2}\left( \Delta \tau \right) ^{2}\left[\frac{2i\omega _{m}^{\ast \ }}{
\hbar }\frac{\delta _{i,j}}{\Delta \tau }+\frac{K_{G}}{\hbar ^{2}}\Re {%
[\gamma _{i}\cdot\gamma _{j}^{\ast }]} \right],\qquad \left[ \mathbf{A}%
_{2}^{\left( N\right) }\right] _{ij}\ =\omega _{m}^{\ast -2}\left( \Delta
\tau \right) ^{2}\ \left[-\frac{2i\omega _{m}^{\ast \ }}{\hbar }\frac{
\delta _{i,j}}{\Delta \tau }+\frac{K_{G}}{\hbar ^{2}}\Re{[\gamma
_{i}\cdot\gamma _{j}^{\ast }]} \right],
\end{equation*}
\begin{equation*}
\left[ \mathbf{J}_{1}^{\left( N\right) }\right] _{i}=\frac{1}{\hbar }\sqrt{
K_{G}}\;\frac{\Delta\tau}{\omega _{m}^{\ast}}\left[ \left( \beta ^{\ast }+
\widetilde{\beta }^{\ast }\right) \gamma _{i}+2\alpha \gamma _{i}^{\ast
}+\Gamma^{\mathcal{N}_{A}\widetilde{\mathcal{N}}_{A}} _{+}\right],\qquad %
\left[ \mathbf{J}_{2}^{\left( N\right) } \right] _{i}=\frac{1}{\hbar }\sqrt{%
K_{G}}\;\frac{\Delta\tau}{\omega _{m}^{\ast}}\left[ \left( \beta ^{\ast }-%
\widetilde{\beta }^{\ast}\right) \gamma_{i}+\Gamma_{-}^{\mathcal{N}_{A}%
\widetilde{\mathcal{N}}_{A}}\right];
\end{equation*}
\end{widetext}
here with $\Re{[x]}$ we indicate the real part of $x$ and
\begin{equation*}
\Gamma_{+}^{\mathcal{N}_{A}\widetilde{\mathcal{N}}_{A}}=\Gamma(t)+\widetilde{%
\Gamma}(t)\qquad\mbox{and}\qquad \Gamma_{-}^{\mathcal{N}_{A}\widetilde{%
\mathcal{N}}_{A}}=\Gamma(t)-\widetilde{\Gamma}(t).
\end{equation*}
We take now the continuous limit as
\begin{widetext}
\begin{eqnarray*}
\omega_m^{*2}\,\mathbf{A}_{1[2]}^{\left( N\right)}
  &\underset{_{\substack{ N\rightarrow \infty
\\ \Delta \tau \rightarrow 0}}} \longrightarrow & \qquad
\mathbf{A}_{1[2]}\left( \tau ^{\prime },\tau ^{\prime \prime
}\right), \\
\frac{\mathbf{J}_{1}^{\left( N\right)}}{\Delta
\tau}&\underset{_{\substack{N\rightarrow \infty  \\ \Delta
\tau\rightarrow 0}}}\longrightarrow &\qquad\mathbf{ J}_{1}(\tau
)=\qquad \frac{1}{\hbar\omega _{m}^{\ast}}\sqrt{K_{G}} \left[
(\beta^{\ast}+\widetilde{\beta }^{\ast }) \gamma(\tau)+2\alpha
\gamma^{\ast}(\tau)+\Gamma_{+}^{\mathcal{N}_{A}\widetilde{\mathcal{N}}
_{A}}(\tau)\right], \\
\frac{\mathbf{J}_{2}^{(N)}}{\Delta \tau }&\underset{_{\substack{
N\rightarrow \infty  \\ \Delta \tau\rightarrow 0}}}\longrightarrow
&\qquad\mathbf{J} _{2}(\tau) =\qquad \frac{1}{\hbar\omega
_{m}^{\ast}}\sqrt{K_{G}}\left[ (\beta ^{\ast }-\widetilde{\beta
}^{\ast }) \gamma(\tau)+\Gamma_{-}^{
\mathcal{N}_{A}\widetilde{\mathcal{N}}_{A}}(\tau) \right],
\end{eqnarray*}
\end{widetext}
\begin{equation*}
\Lambda\equiv\lim_{\substack{ N\rightarrow \infty  \\ \Delta
\tau\rightarrow 0}} \frac{(2\pi)^{N}}{c^{2}\sqrt{\det
\mathbf{A}_{1}\det \mathbf{A} _{2}}},\quad \mbox{with}\quad
c^{2}=\frac{2\pi \hbar \omega _{m}^{\ast}}{\Delta \tau }.
\end{equation*}
Alternatively, the function $\Lambda(t)$ can also be obtained from
the normalization condition $
\langle\langle\Psi(t)||\Psi(t)\rangle\rangle=1.$

Defining now, for general functions $f(\tau )$ and $g(\tau )$,
\begin{equation}\label{eqn:f1}
  \mathcal{F}_{1[2]}^{f,g}\left( \tau \right) =\int\limits_{0}^{\tau
}\int\limits_{0}^{\tau }d\tau ^{\prime }d\tau ^{\prime \prime
}f\left( \tau ^{\prime }\right) \mathbf{A}_{1[2]}^{-1}\left( \tau
^{\prime },\tau ^{\prime \prime }\right) g\left( \tau ^{\prime
\prime }\right) ,
\end{equation}
we get
\begin{widetext}
\begin{eqnarray*}
&&K^{\mathcal{N}_{A}\widetilde{\mathcal{N}}_{A}}(\beta ,\widetilde{\beta }%
;t)= \\
&=&\Lambda \left( t\right) \;e^{-\frac{|\beta |^{2}}{2}-\frac{|\alpha |^{2}}{%
2}+\beta ^{\ast }\alpha }e^{-\frac{|\widetilde{\beta }|^{2}}{2}-\frac{%
|\alpha |^{2}}{2}+\widetilde{\beta }^{\ast }\alpha }\times  \\
&\times &e^{-\frac{1}{2\hbar ^{2}}K_{G}\int\limits_{0}^{\omega _{m}^{\ast
}t}\int\limits_{0}^{\omega _{m}^{\ast }t}d\tau ^{\prime }d\tau ^{\prime
\prime }\left[ \left( \beta ^{\ast }+\widetilde{\beta }^{\ast }\right)
\gamma \left( \tau ^{\prime }\right) +2\alpha \gamma ^{\ast }\left( \tau
^{\prime }\right) +\Gamma _{+}^{\mathcal{N}_{A}\widetilde{\mathcal{N}}%
_{A}}\left( \tau ^{\prime }\right) \right]
\mathbf{A}_{1}^{-1}\left( \tau
^{\prime },\tau ^{\prime \prime }\right) \left[ \left( \beta ^{\ast }+%
\widetilde{\beta }^{\ast }\right) \gamma \left( \tau ^{\prime \prime
}\right) +2\alpha \gamma ^{\ast }\left( \tau ^{\prime \prime }\right)
+\Gamma _{+}^{\mathcal{N}_{A}\widetilde{\mathcal{N}}_{A}}\left( \tau
^{\prime \prime }\right) \right] }\times  \\
&\times &e^{-\frac{1}{2\hbar ^{2}}K_{G}\int\limits_{0}^{\omega _{m}^{\ast
}t}\int\limits_{0}^{\omega _{m}^{\ast }t}d\tau ^{\prime }d\tau ^{\prime
\prime }\left[ \left( \beta ^{\ast }-\widetilde{\beta }^{\ast }\right)
\gamma \left( \tau ^{\prime }\right) +\Gamma _{-}^{\mathcal{N}_{A}\widetilde{%
\mathcal{N}}_{A}}\left( \tau ^{\prime }\right) \right] \mathbf{A}%
_{2}^{-1}\left( \tau ^{\prime },\tau ^{\prime \prime }\right) \left[ \left(
\beta ^{\ast }-\widetilde{\beta }^{\ast }\right) \gamma \left( \tau ^{\prime
\prime }\right) +\Gamma _{-}^{\mathcal{N}_{A}\widetilde{\mathcal{N}}%
_{A}}\left( \tau ^{\prime \prime }\right) \right] }=
\end{eqnarray*}
\begin{eqnarray*}
&=&\Lambda (t)\;e^{-\frac{|\beta |^{2}}{2}-\frac{|\alpha |^{2}}{2}+\beta
^{\ast }\alpha }e^{-\frac{|\widetilde{\beta }|^{2}}{2}-\frac{|\alpha |^{2}}{2%
}+\widetilde{\beta }^{\ast }\alpha }\times  \\
&\times &e^{-\frac{1}{2\hbar ^{2}}K_{G}\left\{ \left( \beta ^{\ast }+%
\widetilde{\beta }^{\ast }\right) ^{2}\int\limits_{0}^{\omega _{m}^{\ast
}t}\int\limits_{0}^{\omega _{m}^{\ast }t}d\tau ^{\prime }d\tau ^{\prime
\prime }\gamma \left( \tau ^{\prime }\right) \mathbf{A}_{1}^{-1}\left( \tau
^{\prime },\tau ^{\prime \prime }\right) \gamma \left( \tau ^{\prime \prime
}\right) +4\alpha \left( \beta ^{\ast }+\widetilde{\beta }^{\ast }\right)
\int\limits_{0}^{\omega _{m}^{\ast \ }t}\int\limits_{0}^{\omega _{m}^{\ast \
}t}d\tau ^{\prime }d\tau ^{\prime \prime }\gamma ^{\ast }\left( \tau
^{\prime }\right) \mathbf{A}_{1}^{-1}\left( \tau ^{\prime },\tau ^{\prime
\prime }\right) \gamma (\tau ^{\prime \prime })\right\} }\times  \\
&\times &e^{-\frac{1}{2\hbar ^{2}}K_{G}\left\{ 2\left( \beta ^{\ast }+%
\widetilde{\beta }^{\ast }\right) \int\limits_{0}^{\omega _{m}^{\ast
}t}\int\limits_{0}^{\omega _{m}^{\ast }t}d\tau ^{\prime }d\tau ^{\prime
\prime }\gamma (\tau ^{\prime })\ \mathbf{A}_{1}^{-1}(\tau ^{\prime },\tau
^{\prime \prime })\Gamma _{+}^{\mathcal{N}_{A}\widetilde{\mathcal{N}}%
_{A}}(\tau ^{\prime \prime })\right\} }\times  \\
&\times &e^{-\frac{1}{2\hbar ^{2}}K_{G}\left\{ \left( \beta ^{\ast }-%
\widetilde{\beta }^{\ast }\right) ^{2}\int\limits_{0}^{\omega _{m}^{\ast
}t}\int\limits_{0}^{\omega _{m}^{\ast }t}d\tau ^{\prime }d\tau ^{\prime
\prime }\gamma \left( \tau ^{\prime }\right) \mathbf{A}_{2}^{-1}\left( \tau
^{\prime },\tau ^{\prime \prime }\right) \gamma \left( \tau ^{\prime \prime
}\right) +2\left( \beta ^{\ast }-\widetilde{\beta }^{\ast }\right)
\int\limits_{0}^{\omega _{m}^{\ast }t}\int\limits_{0}^{\omega _{m}^{\ast
}t}d\tau ^{\prime }d\tau ^{\prime \prime }\gamma \left( \tau ^{\prime
}\right) \ \mathbf{A}_{2}^{-1}\left( \tau ^{\prime },\tau ^{\prime \prime
}\right) \ \Gamma _{-}^{\mathcal{N}_{A}\widetilde{\mathcal{N}}_{A}}\left(
\tau ^{\prime \prime }\right) \right\} }\times  \\
&\times &e^{-\frac{1}{2\hbar ^{2}}K_{G}\left\{ 4\alpha
\int\limits_{0}^{\omega _{m}^{\ast }t}\int\limits_{0}^{\omega _{m}^{\ast
}t}d\tau ^{\prime }d\tau ^{\prime \prime }\gamma ^{\ast }\left( \tau
^{\prime }\right) \ \mathbf{A}_{1}^{-1}\left( \tau ^{\prime },\tau ^{\prime
\prime }\right) \ \Gamma _{+}^{\mathcal{N}_{A}\widetilde{\mathcal{N}}%
_{A}}\left( \tau ^{\prime \prime }\right) \right\} }\times  \\
&\times &e^{-\frac{1}{2\hbar ^{2}}K_{G}\left\{ 4\alpha
^{2}\int\limits_{0}^{\omega _{m}^{\ast }t}\int\limits_{0}^{\omega _{m}^{\ast
}t}d\tau ^{\prime }d\tau ^{\prime \prime }\gamma ^{\ast }\left( \tau
^{\prime }\right) \mathbf{A}_{1}^{-1}\left( \tau ^{\prime },\tau ^{\prime
\prime }\right) \gamma ^{\ast }\left( \tau ^{\prime \prime }\right)
+\int\limits_{0}^{\omega _{m}^{\ast }t}\int\limits_{0}^{\omega _{m}^{\ast
}t}d\tau ^{\prime }d\tau ^{\prime \prime }\Gamma _{+}^{\mathcal{N}_{A}%
\widetilde{\mathcal{N}}_{A}}\left( \tau ^{\prime }\right) \mathbf{A}%
_{1}^{-1}\left( \tau ^{\prime },\tau ^{\prime \prime }\right) \Gamma _{+}^{%
\mathcal{N}_{A}\widetilde{\mathcal{N}}_{A}}\left( \tau ^{\prime \prime
}\right) \right\} }\times  \\
&\times &e^{-\frac{1}{2\hbar ^{2}}K_{G}\left\{ \int\limits_{0}^{\omega
_{m}^{\ast }t}\int\limits_{0}^{\omega _{m}^{\ast }t}d\tau ^{\prime }d\tau
^{\prime \prime }\Gamma _{-}^{\mathcal{N}_{A}\widetilde{\mathcal{N}}%
_{A}}\left( \tau ^{\prime }\right) \mathbf{A}_{2}^{-1}\left( \tau ^{\prime
},\tau ^{\prime \prime }\right) \ \Gamma _{-}^{\mathcal{N}_{A}\widetilde{%
\mathcal{N}}_{A}}\left( \tau ^{\prime \prime }\right) \right\} }.
\end{eqnarray*}
Finally
\begin{equation}
\begin{split}
{K}^{\mathcal{N}_{A}\widetilde{\mathcal{N}}_{A}}(\beta
,\widetilde{ \beta };t)& =\Lambda \left( t\right) e^{-\frac{|\beta
|^{2}}{2}-\frac{| \widetilde{\beta }|^{2}}{2}+\beta ^{\ast }\alpha
+\widetilde{\beta }^{\ast }\alpha -|\alpha
|^{2}}e^{-\frac{1}{2\hbar ^{2}}K_{G}\left\{ 4\alpha \left[
\mathcal{F}_{1}^{\gamma ^{\ast },\Gamma _{+}}+\alpha
\mathcal{F}_{1}^{\gamma ^{\ast },\gamma ^{\ast }}\right]
+\mathcal{F}_{1}^{\Gamma _{+},\Gamma _{+}}+
\mathcal{F}_{2}^{\Gamma _{-},\Gamma _{-}}\right\} }\times  \\
& \times e^{-\frac{1}{2\hbar ^{2}}K_{G}\left\{
\mathcal{F}_{1}^{\gamma ,\gamma }\left( \beta ^{\ast
}+\widetilde{\beta }^{\ast }\right) ^{2}+ \mathcal{F}_{2}^{\gamma
,\gamma }\left( \beta ^{\ast }-\widetilde{\beta } ^{\ast }\right)
^{2}+2\left( \beta ^{\ast }+\widetilde{\beta }^{\ast }\right)
\left[ 2\alpha \mathcal{F}_{1}^{\gamma ^{\ast },\gamma }\
+\mathcal{ F}_{1}^{\gamma ,\Gamma _{+}}\right]
+2\mathcal{F}_{2}^{\gamma ,\Gamma _{-}}\left( \beta ^{\ast
}-\widetilde{\beta }^{\ast }\right) \ \right\} }.
\end{split}
\label{kappa}
\end{equation}
\end{widetext}
Inverse operators $\mathbf{A}_{1}^{-1},\mathbf{A}_{2}^{-1}$ have
been evaluated numerically.

Note that the dependence on $\beta ,\widetilde{\beta }$ is present
only in the second factor (second and third row), while the
dependence on photon states is hidden in $\Gamma
_{+}^{\mathcal{N}_{A} \widetilde{\mathcal{N}}_{A}},\Gamma
_{-}^{\mathcal{N}_{A}\widetilde{\mathcal{ N}}_{A}}$.

Schr\"{o}dinger state at time $t$ is then given by Equation
(\ref{eqn:prsch}).
\bigskip
\section{Integrals of visibility}
\setcounter{equation}{0} Integrals (I) and (II) appearing in
Equation (\ref{visibbility}) are given by
\begin{widetext}
\begin{eqnarray*}
&&(I) =\frac{1}{4\pi^4}e^{i\kappa ^{2}\left( \omega^* _{m}t-\sin \omega^*
_{m}t\right) }\mathbb{K}^{10}\left( t\right) \mathbb{K}^{00\,*}\left(
t\right) \times \\
&\times&\iint d^{2}\beta d^{2}\beta ^{\prime }e^{i\kappa\Im[
\beta(1-e^{-i\omega_m^*t})]}L\left( \beta ,\beta ^{\prime
}\right)h(\beta,\beta^{\prime }) K_{1}^{10}\left( \beta \right)
K_{3}^{00}\left( \beta ^{\prime }\right)
\iint d^{2}\widetilde{\beta } d^{2}\widetilde{\beta } ^{\prime
}K_{2}^{10}( \widetilde{\beta }) K_{4}^{00}( \widetilde{\beta
}^{\prime })g(\widetilde{\beta},\widetilde{\beta }^{\prime })
H_{c}( \widetilde{\beta }, \widetilde{\beta }^{\prime }),
\\
&&\left( II\right) =\frac{1}{4\pi^4}e^{i\kappa ^{2}\left( \omega^*
_{m}t-\sin \omega^* _{m}t\right) }\mathbb{K}^{11}\left( t\right)
\mathbb{K}
^{01\,*}\left( t\right)\times \\
&\times& \iint d^{2}\beta d^{2}\beta ^{\prime }e^{i\kappa\Im[
\beta(1-e^{-i\omega_m^*t})]}L\left( \beta ,\beta ^{\prime
}\right)h(\beta,\beta^{\prime }) K_{1}^{11}\left( \beta \right)
K_{3}^{01}\left( \beta ^{\prime }\right)
\iint d^{2}\widetilde{\beta } d^{2} \widetilde{\beta }^{\prime
}K_{2}^{11}( \widetilde{\beta }) K_{4}^{01}( \widetilde{\beta
}^{\prime })\varepsilon(\widetilde{\beta}
)\delta(\widetilde{\beta}^{\prime
})g(\widetilde{\beta},\widetilde{\beta} ^{\prime }) H_{l}(
\widetilde{ \beta },\widetilde{\beta }^{\prime }).
\end{eqnarray*}
\end{widetext}


\section{Calculation of the Wigner Function}
\setcounter{equation}{0} In the following we present the details of the calculation of the Wigner function.
Let's start by computing $\rho _{m}\left( t\right).$ Defining
\begin{widetext}
\begin{eqnarray*}
\left\vert \left\vert \Psi \left( t\right) \right\rangle
\right\rangle_{\varphi } &=&\left\langle \varphi \right\vert
\otimes \left\langle \varphi \right\vert \left\vert \left\vert
\Psi \left( t\right) \right\rangle
\right\rangle = \\
&=&\frac{1}{2\pi ^{2}}\iint_{-\infty }^{\infty }d^{2}\beta d^{2}\widetilde{%
\beta }\biggl[K^{00}(\beta ,\widetilde{\beta })\left\vert \beta
_{c}\right\rangle \left\vert \widetilde{\beta }_{c}\right\rangle
+K^{01}(\beta ,\widetilde{\beta })f(\widetilde{\beta })e^{-i\theta
}\left\vert \beta _{c}\right\rangle \left\vert \widetilde{\beta }%
_{l}\right\rangle + \\
&+&K^{10}(\beta ,\widetilde{\beta })f(\beta )e^{-i\theta }\left\vert \beta
_{l}\right\rangle \left\vert \widetilde{\beta }_{c}\right\rangle
+K^{11}(\beta ,\widetilde{\beta })f(\beta )f(\widetilde{\beta })e^{-2i\theta
}\left\vert \beta _{l}\right\rangle \left\vert \widetilde{\beta }%
_{l}\right\rangle \biggr]
\end{eqnarray*}
\end{widetext}
and
$$\rho_\varphi(t)=\|\Psi(t)\rangle\rangle_{\varphi}\langle\langle\Psi(t)\|,$$
the density matrix of the physical mirror is:
\begin{widetext}
\begin{eqnarray*}
\rho _{m}\left( t\right)&=& Tr(\rho _{\varphi }\left( t\right)
)=\frac{1}{ \pi }\int d^{2}\widetilde{\chi }\left\langle
\widetilde{\chi }\right\vert \rho _{\varphi }\left\vert
\widetilde{\chi }\right\rangle = \\
&=&\frac{1}{4\pi ^{4}}\int d^{2}(\beta ,\widetilde{\beta },\beta
^{\prime }, \widetilde{\beta ^{\prime }})\biggl(\alpha
_{1}\left\vert \beta _{c}\right\rangle \left\langle \beta
_{c}^{\prime }\right\vert +\alpha _{2}\left\vert \beta
_{l}\right\rangle \left\langle \beta _{l}^{\prime }\right\vert +
\alpha _{3}\left\vert \beta _{c}\right\rangle \left\langle \beta
_{l}^{\prime }\right\vert +\alpha _{4}\left\vert \beta
_{l}\right\rangle \left\langle \beta _{c}^{\prime }\right\vert
\biggr),
\end{eqnarray*}
\end{widetext}
where
\begin{widetext}
\begin{eqnarray*}
\alpha _{1}(\beta ,\widetilde{\beta },\beta ^{\prime },\widetilde{\beta }%
^{\prime }) &=&K^{00}(\beta ,\widetilde{\beta })K^{\ast 00}(\beta ^{\prime },%
\widetilde{\beta }^{\prime})e^{ \widetilde{\beta }_{c}\widetilde{\beta }%
_{c}^{\prime \ast }-\frac{1}{2}\left\vert \widetilde{\beta }_{c}\right\vert
^{2}-\frac{1}{2}\left\vert \widetilde{\beta }_{c}^{\prime }\right\vert ^{2}}+
\\
&+&K^{00}(\beta ,\widetilde{\beta })K^{\ast 01}(\beta ^{\prime },\widetilde{%
\beta ^{\prime }})e^{i\theta }f^{\ast }(\widetilde{\beta
}^{\prime})e^{
\widetilde{\beta }_{c}\widetilde{\beta _{l}^{\prime }}^{\ast }-\frac{1}{2}%
\left\vert \widetilde{\beta }_{c}\right\vert ^{2}-\frac{1}{2}\left\vert
\widetilde{\beta _{l}^{\prime }}\right\vert ^{2}}+ \\
&+&K^{01}(\beta ,\widetilde{\beta })K^{\ast 00}(\beta ^{\prime },\widetilde{%
\beta ^{\prime }})e^{-i\theta }f(\widetilde{\beta })\,e^{\widetilde{\beta }%
_{l}\widetilde{\beta }_{c}^{\prime ^{\ast }}-\frac{1}{2}\left\vert
\widetilde{\beta }_{l}\right\vert ^{2}-\frac{1}{2}\left\vert \widetilde{%
\beta }_{c}^{\prime }\right\vert ^{2}}+ \\
&+&K^{01}(\beta ,\widetilde{\beta })K^{\ast 01}(\beta ^{\prime },\widetilde{%
\beta ^{\prime }})f(\widetilde{\beta })f^{\ast }(\widetilde{\beta
}^{\prime})e^{
\widetilde{\beta }_{l}\widetilde{\beta _{l}^{\prime }}^{\ast }-\frac{1}{2}%
\left\vert \widetilde{\beta }_{l}\right\vert
^{2}-\frac{1}{2}\left\vert \widetilde{\beta _{l}^{\prime
}}\right\vert ^{2}},
\end{eqnarray*}

\begin{eqnarray*}
\alpha _{2}(\beta ,\widetilde{\beta },\beta ^{\prime },\widetilde{\beta
^{\prime }}) &=&K^{10}(\beta ,\widetilde{\beta })K^{\ast 10}(\beta ^{\prime
},\widetilde{\beta ^{\prime }})f(\beta )f^{\ast }(\beta ^{\prime })\,e^{%
\widetilde{\beta }_{c}\widetilde{\beta }_{c}^{\prime \ast }-\frac{1}{2}%
\left\vert \widetilde{\beta}_{c}\right\vert ^{2}-\frac{1}{2}\left\vert
\widetilde{\beta } _{c}^{\prime}\right\vert ^{2}}+ \\
&+&K^{10}(\beta ,\widetilde{ \beta })K^{\ast 11}(\beta ^{\prime },\widetilde{%
\beta ^{\prime }})f(\beta )f^{\ast }(\beta ^{\prime })e^{i\theta }f^{\ast }(%
\widetilde{\beta} ^{\prime })\,e^{\widetilde{\beta }_{c}\widetilde{ \beta
_{l}^{\prime }}^{\ast }-\frac{1}{2}\left\vert \widetilde{\beta }
_{c}\right\vert ^{2}-\frac{1}{2}\left\vert \widetilde{\beta ^{\prime }}
_{l}\right\vert ^{2}}+ \\
&+&K^{11}(\beta ,\widetilde{\beta })K^{\ast 10}(\beta ^{\prime },\widetilde{
\beta ^{\prime }})e^{-i\theta }f(\beta )f^{\ast }(\beta ^{\prime })f(%
\widetilde{\beta } )\,e^{\widetilde{\beta }_{l} \widetilde{\beta }%
_{c}^{\prime^*} -\frac{1}{2}\left\vert \widetilde{ \beta }_{l}\right\vert
^{2}-\frac{1}{2}\left\vert \widetilde{\beta} ^{\prime } _{c}\right\vert
^{2}}+ \\
&+&K^{11}(\beta ,\widetilde{\beta })K^{\ast 11}(\beta ^{\prime }, \widetilde{%
\beta ^{\prime }})f(\beta)f(\widetilde{\beta})f^{\ast }(\beta
^{\prime })f^{\ast }(\widetilde{\beta}^{\prime})e^{
\widetilde{\beta }_{l}\widetilde{\beta
_{l}^{\prime }}^{\ast }-\frac{1}{2 }\left\vert \widetilde{\beta }%
_{l}\right\vert ^{2}-\frac{1}{2}\left\vert \widetilde{\beta ^{\prime }_{l}}%
\right\vert ^{2}},
\end{eqnarray*}

\begin{eqnarray*}
\alpha _{3}(\beta ,\widetilde{\beta },\beta ^{\prime },\widetilde{\beta
^{\prime }}) &=&K^{00}(\beta ,\widetilde{\beta })K^{\ast 10}(\beta ^{\prime
},\widetilde{\beta ^{\prime }})e^{i\theta }f^{\ast }(\beta ^{\prime })\,e^{
\widetilde{\beta }_{c}\widetilde{\beta} _{c}^{\prime^*}-\frac{1}{2}
\left\vert \widetilde{\beta }_{c}\right\vert ^{2}-\frac{1}{2}\left\vert
\widetilde{\beta }^{\prime }_{c}\right\vert ^{2}}+ \\
&+&K^{00}(\beta ,\widetilde{ \beta })K^{\ast 11}(\beta ^{\prime },\widetilde{%
\beta ^{\prime }})e^{2i\theta }f^{\ast }(\beta ^{\prime })f^{\ast }(%
\widetilde{\beta }^{\prime })\,e^{ \widetilde{\beta }_{c}\widetilde{\beta
_{l}^{\prime }}^{\ast }-\frac{1}{2} \left\vert \widetilde{\beta }%
_{c}\right\vert ^{2}-\frac{1}{2}\left\vert \widetilde{\beta ^{\prime }_{l}}%
\right\vert ^{2}}+ \\
&+&K^{01}(\beta ,\widetilde{\beta })K^{\ast 10}(\beta ^{\prime },\widetilde{
\beta ^{\prime }})f(\widetilde{\beta })f^{\ast }(\beta ^{\prime })\,e^{%
\widetilde{\beta }_{l} \widetilde{\beta }_{c}^{\prime^*} -\frac{1}{2}%
\left\vert \widetilde{ \beta }_{l}\right\vert ^{2}-\frac{1}{2}\left\vert
\widetilde{\beta} ^{\prime } _{c}\right\vert ^{2}}+ \\
&+&K^{01}(\beta ,\widetilde{\beta })K^{\ast 11}(\beta ^{\prime }, \widetilde{%
\beta ^{\prime }})e^{i\theta }f(\widetilde{\beta })f^{\ast }(\beta ^{\prime
})f^{\ast }(\widetilde{\beta ^{\prime }})e^{\widetilde{\beta }_{l}\widetilde{%
\beta _{l}^{\prime }}^{\ast }-\frac{1}{2 }\left\vert \widetilde{\beta }%
_{l}\right\vert ^{2}-\frac{1}{2}\left\vert \widetilde{\beta ^{\prime }_{l}}%
\right\vert ^{2}},
\end{eqnarray*}

\begin{eqnarray*}
\alpha _{4}(\beta ,\widetilde{\beta },\beta ^{\prime },\widetilde{\beta
^{\prime }}) &=&K^{10}(\beta ,\widetilde{\beta })K^{\ast 00}(\beta ^{\prime
},\widetilde{\beta ^{\prime }})e^{-i\theta }f(\beta )\,e^{\widetilde{\beta }%
_{c}\widetilde{\beta }_{c}^{\prime \ast }-\frac{1}{2}\left\vert \widetilde{%
\beta}_{c}\right\vert ^{2}-\frac{1}{2}\left\vert \widetilde{\beta }
_{c}^{\prime}\right\vert ^{2}}+ \\
&+&K^{10}(\beta ,\widetilde{ \beta })K^{\ast 01}(\beta ^{\prime },\widetilde{%
\beta ^{\prime }})f(\beta )f^{\ast }(\widetilde{\beta} ^{\prime })\,e^{%
\widetilde{\beta }_{c}\widetilde{ \beta _{l}^{\prime }}^{\ast }-\frac{1}{2}%
\left\vert \widetilde{\beta } _{c}\right\vert ^{2}-\frac{1}{2}\left\vert
\widetilde{\beta ^{\prime }} _{l}\right\vert ^{2}}+ \\
&+&K^{11}(\beta ,\widetilde{\beta })K^{\ast 00}(\beta ^{\prime },\widetilde{
\beta ^{\prime }})e^{-2i\theta }f(\beta )f(\widetilde{\beta } )\,e^{%
\widetilde{\beta }_{l} \widetilde{\beta }_{c}^{\prime^*} -\frac{1}{2}%
\left\vert \widetilde{ \beta }_{l}\right\vert ^{2}-\frac{1}{2}\left\vert
\widetilde{\beta} ^{\prime } _{c}\right\vert ^{2}}+ \\
&+&K^{11}(\beta ,\widetilde{\beta })K^{\ast 01}(\beta ^{\prime }, \widetilde{%
\beta ^{\prime }})e^{-i\theta }f(\beta)f(\widetilde{\beta})f^{\ast
}( \widetilde{\beta}^{\prime })e^{\widetilde{\beta
}_{l}\widetilde{\beta _{l}^{\prime }}^{\ast }-\frac{1}{2
}\left\vert \widetilde{\beta } _{l}\right\vert
^{2}-\frac{1}{2}\left\vert \widetilde{\beta ^{\prime }_{l}}
\right\vert ^{2}}.
\end{eqnarray*}

By taking the trace

\begin{eqnarray*}
Tr\left[ \rho _{m}\left( t\right) \ e^{\lambda b^{\dagger
}-\lambda ^{\ast }b}\right]  &=&\frac{1}{\pi }\int d^{2}\chi
\left\langle \chi \right\vert \rho _{m}\left( t\right) e^{\lambda
b^{\dagger }-\lambda ^{\ast
}b}\left\vert \chi \right\rangle = \\
&=&\frac{1}{4\pi ^{5}}\int d^{2}(\beta ,\widetilde{\beta },\beta ^{\prime },%
\widetilde{\beta ^{\prime }})\int d^{2}\chi e^{-\left\vert \chi \right\vert
^{2}-\lambda ^{\ast }\chi -\frac{1}{2}\left\vert \lambda \right\vert ^{2}}%
\biggl(\alpha _{1}e^{\beta _{c}\chi ^{\ast }-\frac{1}{2}\left\vert \beta
_{c}\right\vert ^{2}-\frac{1}{2}\left\vert \beta _{c}^{\prime }\right\vert
^{2}+\beta _{c}^{\prime ^{\ast }}\chi +\beta _{c}^{\prime ^{\ast }}\lambda }+
\\
&+&\alpha _{2}e^{\beta _{l}\chi ^{\ast }-\frac{1}{2}\left\vert \beta
_{l}\right\vert ^{2}-\frac{1}{2}\left\vert \beta _{l}^{\prime }\right\vert
^{2}+\beta _{l}^{\prime ^{\ast }}\chi +\beta _{l}^{\prime ^{\ast }}\lambda
}+\alpha _{3}e^{\beta _{c}\chi ^{\ast }-\frac{1}{2}\left\vert \beta
_{c}\right\vert ^{2}-\frac{1}{2}\left\vert \beta _{l}^{\prime }\right\vert
^{2}+\beta _{l}^{\prime ^{\ast }}\chi +\beta _{l}^{\prime ^{\ast }}\lambda }+
\\
&+&\alpha _{4}e^{\beta _{l}\chi ^{\ast }-\frac{1}{2}\left\vert \beta
_{l}\right\vert ^{2}-\frac{1}{2}\left\vert \beta _{c}^{\prime }\right\vert
^{2}+\beta _{c}^{\prime ^{\ast }}\chi +\beta _{c}^{\prime ^{\ast }}\lambda }%
\biggr)
\end{eqnarray*}

we finally obtain Equation (\ref{Wigner_int}).

\end{widetext}

\end{document}